\documentclass[aps,pra,twocolumn,showpacs,superscriptaddress]{revtex4}
\usepackage{graphicx}
\usepackage{amsmath}
\usepackage{amssymb}

\input{epsf}
\begin{document}

\title{
Few-body resonances of unequal-mass systems
with infinite interspecies two-body $s$-wave scattering length}

\author{D. Blume and K. M. Daily}
\affiliation{Department of Physics and Astronomy,
Washington State University,
  Pullman, Washington 99164-2814, USA}

\date{\today}

\begin{abstract}
Two-component Fermi and Bose gases with infinitely
large interspecies $s$-wave scattering length $a_s$ exhibit a
variety of intriguing properties. Among these
are the scale invariance of two-component Fermi gases
with equal masses, and the favorable scaling
of Efimov features for two-component Bose gases and Bose-Fermi
mixtures with unequal
masses.
This paper builds on our
earlier work
[D. Blume and K. M. Daily, arXiv:1006.5002] 
and presents a detailed 
discussion of our studies of small unequal-mass two-component 
systems 
with infinite $a_s$ in the regime where 
three-body Efimov physics is absent.
We report on non-universal few-body resonances.
Just like with two-body systems on resonance, few-body systems 
have a zero-energy bound state in free space
and a diverging generalized scattering length.
Our calculations are performed
within a non-perturbative microscopic framework
and
investigate
the energetics and structural properties of small unequal-mass
two-component systems
as functions of the mass ratio
$\kappa$,
and the numbers $N_{1}$ and $N_2$
of heavy and light atoms.
For purely attractive Gaussian two-body interactions,
we find that 
the $(N_1,N_2)=(2,1)$ and $(3,1)$ systems exhibit 
three-body and four-body resonances at mass ratios 
$\kappa = 12.314(2)$
and $10.4(2)$, respectively.
The 
three- and four-particle systems on resonance are found to 
be large. This suggests that the corresponding wave function
has relatively small overlap with deeply-bound dimers, trimers or
larger clusters and that the three- and four-body systems on resonance 
have a comparatively long lifetime.
Thus, it seems feasible that the features discussed in this paper
can be probed experimentally with present-day technology.
\end{abstract}

\pacs{03.75.Ss,05.30.Fk,34.50.-s}

\maketitle

\section{Introduction}
\label{sec_introduction}
Resonances arise in many different branches of physics.
Near a resonance, certain physical observables are
strongly enhanced, allowing one to probe the
underlying physics at its ``extreme''.
Resonances come in various types such as 
tunneling resonances (see, e.g., Ref.~\cite{ricc84}),
parametric resonances (see, e.g., Ref.~\cite{kofm94}), 
stochastic resonances (see, e.g., Ref.~\cite{gamm98}),
or Fano-Feshbach
resonances (see, e.g., Ref.~\cite{chin10}).
Two-body Fano-Feshbach resonances
play a key role in the study
of ultracold atomic gases as they allow for
the tuning of the interspecies and intraspecies
$s$-wave scattering length $a_s$ essentially
at will while leaving all other system parameters
essentially unchanged~\cite{stwa76,ties93,inou98,corn00}. 
In the ultracold regime, the scattering lengths determine 
the effective interaction strengths of the system.
Thus, the tunability of $a_s$ allows for the realization of
effectively repulsive and effectively attractive systems.
Furthermore, the interaction strength can be tuned such that the
system is effectively non-interacting
or infinitely strongly interacting.
Recent experiments 
that rely on this tunability include the study of the
BCS-BEC crossover problem~\cite{grei03,zwie03,bour03,stre03,rega04,zwie04,kina04}, 
the polaron 
problem~\cite{zwie09}, 
and the creation of bosonic and fermionic Feshbach 
molecules~\cite{stre03,xu03,joch03,cubi03,durr04}.
These Feshbach molecules form an ideal starting point
for creating polar molecules  in the rovibrational ground 
state~\cite{ye,grimm,weid}.

Fano-Feshbach resonances have also played a key role in recent
experiments aimed at studying few-body 
physics.
The most prominent example is the observation of
Efimov 
physics~\cite{efimov1,efimov2,braatenreview,krae06,knoo09,ferl09,zacc09,poll09},
which relies on tuning the 
$s$-wave scattering length over several orders of magnitude.
Efimov physics manifests itself most clearly
in the vicinity of resonances. In particular, Efimov physics
in the bosonic three-particle sector has been studied through the 
observation of enhanced three-body recombination loss
rates near atom-dimer and atom-atom-atom 
resonances~\cite{krae06,knoo09}.
An atom-dimer resonance exists when the
binding energy of the Efimov trimer equals that of the dimer.
An atom-atom-atom resonance exists when the Efimov trimer
has a zero-energy binding energy, i.e., sits right at
threshold.
Similarly, in the bosonic
four-particle sector, the expected scaling
associated with Efimov physics has been confirmed through 
measurements of enhanced four-body
recombination loss rates at atom-atom-atom-atom,
atom-trimer and dimer-dimer 
resonances~\cite{ferl09,zacc09,poll09,stec09a,dinc09}.
An intriguing aspect of the features associated with Efimov physics
is that the characterization of the three- and four-body sectors
requires, for one-component Bose gases, just two parameters, the 
$s$-wave scattering length $a_s$ and a three-body parameter, which can
for example be parametrized in
terms of the binding energy of one of the
Efimov trimers~\cite{stec09a,hammer}. These two parameters determine the 
positions of all universal resonances eluded to above.
Efimov physics can also dominate
the behavior
of fermionic systems consisting of
two or more components and of Bose-Fermi mixtures
if the system
parameters are tuned 
appropriately~\cite{petrov,esry1,esry2,otte08,wenz09,huck09,will09}.

This paper investigates few-body resonances for 
two-component systems with infinitely large 
interspecies $s$-wave scattering length $a_s$ in the regime where
three-body Efimov physics is absent.
Building on earlier 
work~\cite{petrov,esry1,esry2,castin,stec07a,nishida,geze09,russian,ritt10,blum10,gand10}, 
we address the following questions:
(Q1) Under which conditions do atom-atom-atom resonances 
occur for the $(2,1)$ system?
(Q2) Under which conditions do atom-atom-atom-atom
resonances occur for the $(3,1)$ system?
(Q3) If the resonances discussed in Q1 and Q2 exist,
what are their
characteristics?
(Q4) What are the differences and commonalities of
system properties derived for finite-range (FR) and zero-range 
(ZR) interactions?
(Q5) What are the implications of our theoretical studies for
experiment~\cite{experimentmass}?

The remainder of this paper is organized as follows.
Section~\ref{sec_hamiltonian} introduces the system Hamiltonian
while
Secs.~\ref{sec_hyperspherical}
and \ref{sec_sv} discuss two complementary approaches for
solving the time-independent Schr\"odinger equation
for small two-component systems with unequal masses.
Section~\ref{sec_results} presents our results for the
energetics and selected structural properties for
various parameter combinations. 
Lastly, Section~\ref{sec_conclusion} summarizes
our main results and concludes.
A discussion of the main ideas and results of our work 
can be found in Ref.~\cite{blum10}. The present paper elaborates on
the theoretical framework and provides a more detailed discussion
of the results and their implications. In addition,
the present paper presents 
structural properties, detailed
comparisons between observables derived within 
the numerical and analytical
frameworks, and results for the $(3,2)$ and $(4,1)$ systems.

\section{Theoretical background}
\label{sec_theory}
This section introduces and
discusses the system Hamiltonian that
underlies our studies, and the techniques employed to
solve the time-independent Schr\"odinger equation
associated with this Hamiltonian. 
Our solutions to the Schr\"odinger equation 
are obtained following two distinctly different
approaches. On the one hand, we pursue an analytical 
treatment that employs hyperspherical coordinates
and determines the solution in terms of some unknown.
On the other hand, we solve the Schr\"odinger equation 
numerically. The numerical solutions are then used to determine 
some of the unknowns that arise in the first approach,
thereby providing an interpretation of the numerical results
and, conversely,
a check of the analytical framework.

\subsection{System Hamiltonian}
\label{sec_hamiltonian}
We consider two-component systems with $N$ particles,
where $N=N_1 + N_2$, under external spherically symmetric 
harmonic
confinement with angular trapping frequency $\omega$.
If $N_i=1$ ($i=1$ or $2$), no permutation symmetry of the $i$th species needs
to be imposed; this implies that the single particle species can
be fermionic or bosonic.
In contrast, if $N_i$ 
is greater than 1, the results depend on the permutation symmetry.
Throughout, we impose fermionic symmetry for systems with $N_i>1$. 
The masses of the two species are denoted
by $m_1$ and $m_2$, and
our model Hamiltonian $H$ 
reads
\begin{eqnarray}
\label{eq_ham}
H= 
\sum_{j=1}^{N_1} \left( \frac{-\hbar^2}{2 m_1} \nabla_{\vec{r}_j}^2 +
\frac{1}{2}m_1 \omega^2 \vec{r}_j^2 \right) + \nonumber \\
\sum_{j=N_1+1}^{N} \left( \frac{-\hbar^2}{2 m_2} \nabla_{\vec{r}_j}^2 +
\frac{1}{2}m_2 \omega^2 \vec{r}_j^2 \right) + 
\sum_{j=1}^{N_1} \sum_{k=N_1+1}^N V_{\mathrm{tb}}(r_{jk}).
\end{eqnarray}
Here, $\vec{r}_j$
denotes the position vector of 
the $j$th particle measured with respect to the trap
center and
$V_{\mathrm{tb}}(r_{jk})$ 
with $r_{jk}=|\vec{r}_j-\vec{r}_k|$
the interspecies interaction potential.
Throughout, we assume that the particles
of the same species are effectively non-interacting,
i.e., we neglect intraspecies interactions. In the case of 
fermionic species this assumption is fullfilled with high
accuracy for most systems since $s$-wave interactions are
forbidden by symmetry and $p$-wave interactions are 
naturally suppressed~\cite{newtonscattering}. 

Although nature provides us 
with only a finite number of discrete mass ratios $\kappa$,
where
\begin{eqnarray}
\kappa= m_1/m_2,
\end{eqnarray} 
we find it useful to treat $\kappa$
as a continuous variable
to unravel the key physics, i.e., to see how the
physics changes as a particularly
interesting mass ratio is approached. Experimentally, the
effective mass ratio of two-component systems
could be tuned by 
loading the system into an optical lattice~\cite{nishida}.
In addition to the mass ratio, we vary the number of 
heavy and light particles, 
the angular momentum and parity of the state
under consideration, and the two-body interaction $V_{\mathrm{tb}}$.

Our calculations are performed for two
classes of interaction potentials, FR and ZR interactions.
Our FR calculations are performed for a
purely attractive Gaussian potential $V_{\mathrm{g}}$ with range $r_0$
and depth $V_0$ ($V_0>0$),
\begin{eqnarray}
\label{eq_fr}
V_{\mathrm{g}}(r)=-V_0 \exp \left[-\left( \frac{r}{\sqrt{2}r_0} \right)^2
\right].
\end{eqnarray}
While the majority of our FR calculations
considers 
the infinite scattering length
limit, i.e., $1/a_s=0$, we also explore 
how the system behavior
changes as $|a_s|$ is decreased. 
In particular, we pick $r_0$ 
and then adjust the depth $V_0$ so that the
two-body potential has the desired free-space $s$-wave scattering
length $a_s$. 
For positive $a_s$, we restrict ourselves to
potentials that support a single $s$-wave two-body
bound state. For negative $a_s$, we restrict ourselves to
potentials that support no two-body bound state.
Our ZR calculations 
employ the Fermi-Huang pseudopotential
$V_{\mathrm{zr}}$~\cite{ferm34,huan57},
\begin{eqnarray}
\label{eq_zr}
V_{\mathrm{zr}}(r)= \frac{2\pi \hbar^2 a_s}{\mu} \delta(\vec{r})
\frac{\partial }{\partial r}r,
\end{eqnarray}
where
$\mu$ denotes the reduced mass,
\begin{eqnarray}
\mu=m_1m_2/(m_1+m_2).
\end{eqnarray}
Our ZR calculations are restricted to 
unitarity, i.e., to the regime where the $s$-wave
scattering length is infinitely large
and where the two-body system in free space is 
at the verge of supporting a zero-energy bound state.

The system under study is characterized by the following length scales:
the harmonic oscillator length $a_{\mathrm{ho}}$,
\begin{eqnarray}
a_{\mathrm{ho}}=\sqrt{\hbar/(2 \mu \omega)},
\end{eqnarray}
the range $r_0$ of the interaction potential
($r_0=0$ for $V_{\mathrm{zr}}$),
and the $s$-wave scattering length $a_s$.
At unitarity, the $s$-wave scattering length no longer defines a meaningful
length scale.
For sufficiently large $\kappa$,
an additional length scale 
is given by the generalized $N$-body scattering length
(see Sec.~\ref{sec_hyperspherical}).
Our FR calculations are performed in the regime where
$r_0 \ll a_{\mathrm{ho}}$.
In fact, one of the goals of this paper is to quantify how
observables obtained for FR interactions approach those 
determined in the ZR limit. As will be shown in
Sec.~\ref{sec_results},
FR effects can be appreciable for two-component
unequal-mass systems.

Our approaches outlined in Secs.~\ref{sec_hyperspherical}
and \ref{sec_sv}
take advantage of the fact that the center-of-mass degrees of 
freedom $\vec{R}_{\mathrm{cm}}$
separate off. Throughout, we assume that the center-of-mass wave function is
in the ground state, and we label
the solutions $\Psi$ to the time-independent Schr\"odinger equation
in the relative coordinates
by the relative orbital angular momentum $L$
and the relative parity $\Pi$. The corresponding relative eigenenergies
will be denoted by $E$.  
The Hamiltonian given in Eq.~(\ref{eq_ham})
describes the trapped system.
As we discuss in Sec.~\ref{sec_hyperspherical},
the eigenenergies of the trapped $N$-body
system can be expressed in
terms of the generalized scattering length
that characterizes
the corresponding $N$-body scattering problem in free space.
This connection between the trapped and free-space systems, which 
is of course well
known for the two-body problem~\cite{busc98}, provides a great deal 
of insight.

\subsection{Hyperspherical coordinate treatment}
\label{sec_hyperspherical}
\begin{figure}
\vspace*{1.5cm}
\includegraphics[angle=0,width=65mm]{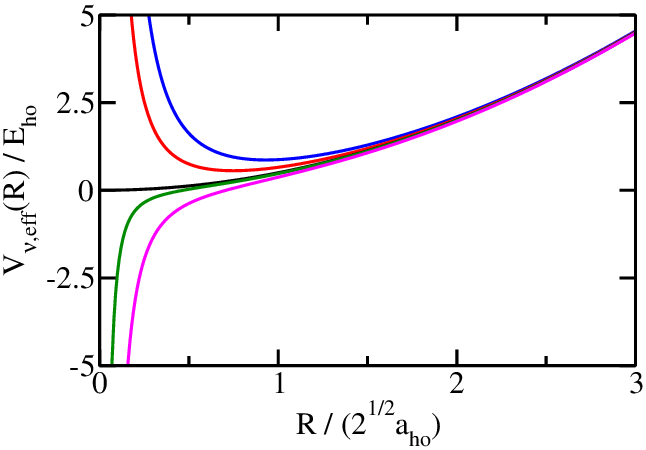}
\vspace*{0.0cm}
\caption{(Color online)
Effective hyperradial potential curve $V_{\nu,{\mathrm{eff}}}(R)$,
Eq.~(\ref{eq_poteff}),
as a function of $R$
for, from top to bottom, $s_{\nu}=1, 3/4, 1/2, 9/20$ and $0$.
}\label{fig_poteff}
\end{figure}

Within the hyperspherical framework~\cite{delv58,delv60,mace68,lin95}, the
$3N-3$ relative coordinates are divided into
$3N-4$ hyperangles, collectively denoted by
$\vec{\Omega}$, and a single length, the hyperradius $R$,
$\mu R^2 = \sum_{j=1}^N m_j (\vec{r}_j-\vec{R}_{\mathrm{cm}})^2$.
In the present context, 
these hyperspherical coordinates are particularly 
appealing since 
the relative wave
function $\Psi$ 
of the unitary system with ZR interactions
has been shown to separate 
for any number of particles into a hyperradial part 
$F_{\nu q}(R)$ and a hyperangular part 
$\Phi_{\nu}(\vec{\Omega})$~\cite{castin},
$\Psi_{\nu q}(R,\vec{\Omega})=R^{-(3N-4)/2} F_{\nu q}(R) 
\Phi_{\nu}(\vec{\Omega})$.
Here, $\nu$ and $q$ denote hyperangular and hyperradial
quantum numbers
for a given angular momentum $L$ and parity $\Pi$; 
$\nu$ takes the values $0,1,\cdots$ while 
$q$ takes non-integer values (see below).
For ZR interactions with $1/a_s=0$,
the relative Schr\"odinger equation can 
therefore be solved in a two-step process.
First, the hyperangular functions $\Phi_{\nu}(\vec{\Omega})$, 
or so-called
channel functions, are determined
by solving the hyperangular Schr\"odinger
equation~\cite{castin}. 
The corresponding eigenvalues are related to the
coefficients
$s_{\nu}$, which
determine the effective potential curves $V_{\nu,{\mathrm{eff}}}(R)$,
\begin{eqnarray}
\label{eq_poteff}
V_{\nu,{\mathrm{eff}}}(R)=\frac{\hbar^2 (s_{\nu}^2-1/4)}{2 \mu R^2}+
\frac{1}{2} \mu \omega^2 R^2.
\end{eqnarray}
Figure~\ref{fig_poteff} shows the effective potential
curves $V_{\nu,{\mathrm{eff}}}(R)$
for different $s_{\nu}$.
For $s_{\nu}=1/2$, 
the hyperangular $1/R^2$ term vanishes
and $V_{\nu,{\mathrm{eff}}}(R)$ reduces to the trapping potential,
i.e., the second term on the right hand side of Eq.~(\ref{eq_poteff}).
For $s_{\nu} > 1/2$ and $s_{\nu}<1/2$, 
the hyperangular $1/R^2$ term is repulsive and attractive, respectively,
and dominates
at small $R$.
Second,
the 
hyperradial Schr\"odinger equation
\begin{eqnarray}
\label{eq_radial}
\left( \frac{-\hbar^2}{2\mu} \frac{\partial^2}{\partial R^2}
+ V_{\nu,{\mathrm{eff}}}(R)
\right) 
F_{\nu q}(R)=
E_{\nu q} F_{\nu q}(R)
\end{eqnarray}
is solved for $F_{\nu q}(R)$ and $E_{\nu q}$.
If one uses FR instead of ZR interactions,
the hyperangular and
hyperradial parts of the wave function $\Psi$,
in general, do not fully separate, implying
non-vanishing coupling matrix elements
between the different hyperangular channel functions. 
In a first approximation, however,
these couplings
can be neglected
if $r_0 \ll a_{\mathrm{ho}}$.
The framework outlined in this section
is thus not only applicable to ZR interaction but also,
at least within an approximative scheme, to
FR interactions (see Refs.~\cite{blum07,stec08}).

\begin{figure}
\vspace*{1.5cm}
\includegraphics[angle=0,width=65mm]{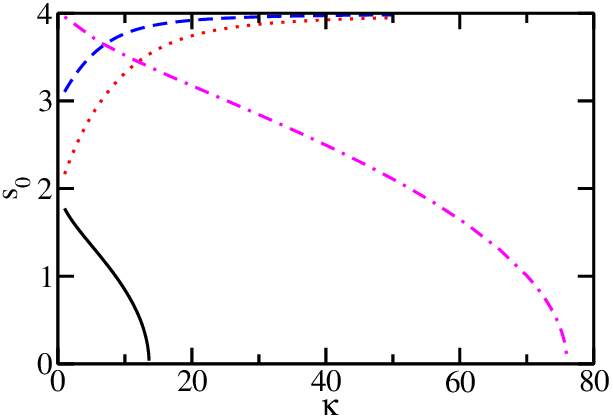}
\vspace*{0.0cm}
\caption{(Color online)
$s_0$ coefficient as a function of $\kappa$ for the $(2,1)$
system at unitarity with ZR interactions and
$L^{\Pi}=0^+$ (dotted line),
$L^{\Pi}=1^-$ (solid line),
$L^{\Pi}=2^+$ (dashed line) and
$L^{\Pi}=3^-$ (dash-dotted line).
Three-body Efimov physics is absent if $s_0>0$, which
corresponds to
$\kappa \lesssim 13.607$ for the $L^{\Pi}=1^-$
system
and to
$\kappa \lesssim 76.0$ for the $L^{\Pi}=3^-$
system.
}\label{fig_s0_n3}
\end{figure}

Solving the 
hyperangular Schr\"odinger equation, i.e., determining the $s_{\nu}$,
is,
in general, a non-trivial task.
For the three-body system with
ZR interactions and infinitely large $a_s$, 
however, the eigenvalues can be 
obtained
by solving a simple transcendental equation
for each $L^{\Pi}$ 
symmetry~\cite{efimov1,efimov2,wern06,russian,ritt10}.
Figure~\ref{fig_s0_n3} shows the $s_0$ coefficients for the
$(2,1)$ system, i.e., the system with
two heavy fermions and one light
atom, at unitarity
with $L=0-3$ and $\Pi=(-1)^L$
as a function of the mass ratio $\kappa$. 
The state with the smallest $s_0$ value has $L^{\Pi}=1^-$ symmetry for
mass ratios $\kappa \lesssim 13.607$
(solid line
in Fig.~\ref{fig_s0_n3}).
The coefficient $s_0$ 
decreases from
$1.773$ for $\kappa=1$ to $0$ for $\kappa \approx 13.607$.
For later reference, we note that $s_0$ equals 1 for $\kappa \approx 8.619$
and $1/2$ for $\kappa=12.313$.
For $\kappa \gtrsim 13.607$, $s_0$ becomes purely imaginary
and 
Efimov physics
comes into play~\cite{efimov1,efimov2,petrov,esry1,esry2}.
The $s_0$ coefficient of other odd $L$ 
states (the dash-dotted line shows the
$s_0$ coefficient for the $L^{\Pi}=3^-$ state)
becomes imaginary for much larger mass ratios.
For even $L$ states, $s_0$ increases with increasing $\kappa$ and approaches
an integer value in the large $\kappa$ limit.
For larger systems
with infinitely large interspecies $s$-wave 
scattering length,
the $s_{\nu}$ coefficients are not known in general.

We now consider the hyperradial 
Schr\"odinger equation, Eq.~(\ref{eq_radial}),
for the $N$-body system,
which
can be solved for arbitrary $s_{\nu}$.
Generalizing the quantum
defect theory type approach from Ref.~\cite{borca}
from $s_{\nu}=1/2$ to arbitrary $s_{\nu}$, 
the solution
$F_{\nu q}(R)$ to the
second order 
differential equation can be written as
\begin{eqnarray}
\label{eq_capfgeneral}
F_{\nu q}(R)=
N_{\nu q} \left[ f_{\nu q}(R)- \tan(\pi \mu_{\nu q}) g_{\nu q}(R) \right],
\end{eqnarray}
where $N_{\nu q}$ denotes a normalization constant.
The quantum defect $\mu_{\nu q}$ determines the 
relative contributions of the regular solution
$f_{\nu q}$ and the irregular solution $g_{\nu q}$,
$f_{\nu q}(x)=A_{\nu q} x^{s_{\nu}+1/2} 
\exp(-x^2/2) _1 F_1(-q,s_{\nu}+1,x^2)$
and
$g_{\nu q}(x)=B_{\nu q} x^{-s_{\nu}+1/2} 
\exp(-x^2/2) _1 F_1(-q-s_{\nu},-s_{\nu}+1,x^2)$.
Here,  $A_{\nu q}$ and $B_{\nu q}$ denote 
constants~\cite{footnoteab},
and $x$ the 
dimensionless hyperradial
coordinate, $x=R/(\sqrt{2}a_{\mathrm{ho}})$.
The non-integer quantum number $q$
is related to the eigenenergy $E_{\nu q}$ through
\begin{eqnarray}
\label{eq_energy}
E_{\nu q}=(2q+s_{\nu}+1)\hbar \omega.
\end{eqnarray}
To determine the allowed values of $\mu_{\nu q}$, 
we enforce that
$F_{\nu q}(x)$ vanishes at large $x$, resulting in
the condition $\sin[\pi (\mu_{\nu q}+q)]=0$.
This condition allows $\mu_{\nu q}$ to be eliminated,
leaving the quantum number $q$ as the only unknown.

The quantization condition, i.e., the allowed $q$ values, 
are determined by investigating the small $x$ behavior
of $F_{\nu q}(x)$.
For $x \rightarrow 0$, $f_{\nu q}(x)$ 
behaves as
$x^{s_{\nu}+1/2}$
and is well-behaved or less strongly diverging than
$x^{-1/2}$ for $s_{\nu}>-1$ while 
$g_{\nu q}(x)$ behaves as $x^{-s_{\nu}+1/2}$ and
diverges faster than $x^{-1/2}$
for $s_{\nu}>1$.
Thus, we consider the regimes $s_{\nu}>1$ and $0 < s_{\nu} < 1$
separately (the Efimov regime with imaginary $s_{\nu}$ is not
treated in this paper).
For $s_{\nu}>1$,
$g_{\nu q}(x)$ must be eliminated~\cite{footnotenorm1}
and $F_{\nu q}(x)$ is determined by the exponentially decaying piece of
$f_{\nu q}(x)$.
The quantization condition becomes,
in agreement with Ref.~\cite{castin}, 
$q=0,1,\cdots$
and
the corresponding energy is referred to as $E_{f,\nu q}$,
\begin{eqnarray}
\label{eq_energyf}
E_{f,\nu q}=(2 q + s_{\nu}+1) \hbar \omega; \; q = 0,1,\cdots.
\end{eqnarray}
For $0 < s_{\nu} < 1$, 
both
$f_{\nu q}(x)$ and $g_{\nu q}(x)$ are well-behaved
and
the allowed $q$ values depend on the
boundary condition 
at small $x$. 
This boundary condition
is
determined by the true atom-atom interactions
and cannot be derived within the ZR framework.
Similarly to the case of Efimov trimers~\cite{efimov1,efimov2,braatenreview},
the value of the
short-range hyperradial boundary condition can be thought of as 
an extra parameter that is needed to specify the solution.
In our case, this parameter characterizes the $N$-body
system and 
can
be parameterized, e.g., by the logarithmic derivative
$L_{\nu q}(x_0)$,
$L_{\nu q}(x_0)=[(\partial F_{\nu q}(x)/\partial x)/F_{\nu q}(x)]_{x=x_0}$.

\begin{figure}
\vspace*{1.5cm}
\includegraphics[angle=0,width=65mm]{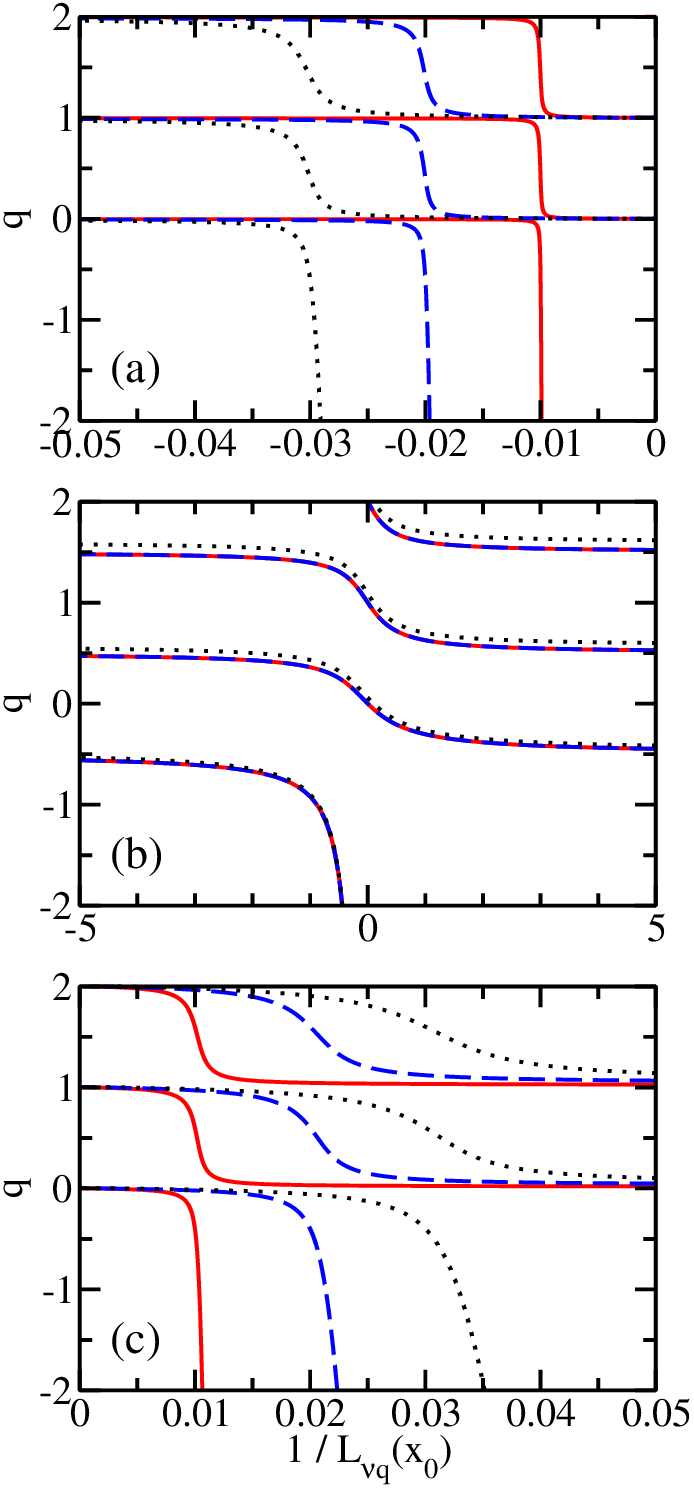}
\vspace*{-0.2cm}
\caption{(Color online)
Non-integer quantum number $q$ as a function of
$[L_{\nu q}(x_0)]^{-1}$ 
for 
(a) $s_{\nu}=3/5$,
(b) $s_{\nu}=1/2$, and
(c) $s_{\nu}=2/5$.
In panels (a) and (c), the solid, dashed and dotted lines
correspond to $x_0=1 \times 10^{-3}$, $2 \times 10^{-3}$ and 
$3 \times 10^{-3}$, respectively.
In panel (b), the solid, dashed and dotted lines
correspond to $x_0=10^{-5}$, $10^{-3}$ and $10^{-1}$, respectively;
the solid and dashed lines are indistinguishable on
the scale shown.
Note that the scale of the $x$-axis is different in all
three panels.
}\label{figLogD}
\end{figure}

For 
$s_{\nu}>0$ ($s_{\nu}$ not equal to an integer),
the normalized hyperradial wave function $F_{\nu q}(R)$
can be compactly written as
\begin{eqnarray}
\label{eq_radial_compact}
F_{\nu q}(x) = \nonumber \\
N_{\nu q}\exp \left(
\frac{-x^2}{2} \right) 
x^{s_{\nu}+1/2} 
U(-q,s_{\nu}+1,x^2),
\end{eqnarray}
where 
 $U$ denotes 
the confluent hypergeometric function of the second kind
and
\begin{eqnarray}
\label{eq_radial_compact_norm}
N_{\nu q} = \sqrt{
\frac{-2 \sin( \pi s_{\nu}) \Gamma(1-q) \Gamma(-q-s_{\nu})}
{\pi + \pi^2 q \cot( \pi q) + q \pi [ \psi(q)-\psi(-q-s_{\nu})]}};
\end{eqnarray}
here, $\psi$ denotes the digamma function~\cite{footnotenorm}.
Using Eq.~(\ref{eq_radial_compact}),
the logarithmic
derivative at $x=x_0$
can be written as
\begin{eqnarray}
\label{eq_logD}
L_{\nu q}(x_0) = \frac{\frac{1}{2}-s_{\nu}}{x_0} - \nonumber \\
x_0 
+ \frac{2 (q+s_{\nu})}{x_0} \frac{U(-q,s_{\nu},x_0^2)}{U(-q,s_{\nu}+1,x_0^2)}.
\end{eqnarray}
Figure~\ref{figLogD}(a) shows the quantum number $q$ for $s_{\nu}=3/5$
for $x_0 \ll 1$ 
(i.e., $x_0=1 \times 10^{-3}$, $2 \times 10^{-3}$ and $3 \times 10^{-3}$)
as a function of the inverse of the logarithmic derivative.
$q$ is approximately equal
to integer values ($q \approx 0,1,\cdots$)
for nearly all values of the logarithmic derivative. 
Exceptions occur for negative $[L_{\nu q}(x_0)]^{-1}$
values with $|1/L_{\nu q}(x_0)|$ small, where $q$ 
drops by roughly 1 over a small range of $[L_{\nu q}(x_0)]^{-1}$
values. The value of $|1/L_{\nu q}(x_0)|$ at which $q$
drops
decreases 
as $x_0$ decreases and reaches $0$ in the ZR limit.
The 
deviation of $q$ from $\approx 0,1,\cdots$ 
signals the existence of an $N$-body resonance,
which arises when
$F_{\nu q}(x)$ is fully determined by the irregular solution 
$g_{\nu q}(x)$.
When only $g_{\nu q}(x)$ contributes, we find that the
quantization condition becomes
$q=-s_{\nu},-s_{\nu}+1,\cdots$
in the ZR limit, in agreement with 
note [43] of Ref.~\cite{castin}.
We refer to the corresponding energies 
as $E_{g,\nu q}$,
\begin{eqnarray}
\label{eq_energyg}
E_{g,\nu q} = (2q+s_{\nu}+1) \hbar \omega ; \; q=-s_{\nu},-s_{\nu}+1,\cdots.
\end{eqnarray}
For $s_{\nu}=3/5$,
$q$ equals $-3/5,2/5,\cdots$
on resonance and $|1/L_{\nu q}(x_0)| = 0$ for $x_0 \rightarrow 0$.

The dependence of $q$ on the inverse logarithmic 
derivative $1/L_{\nu q}(x_0)$ for $s_{\nu}=2/5$ 
[Fig.~\ref{figLogD}(c)] is similar to that for
$s_{\nu}=3/5$ 
[Fig.~\ref{figLogD}(a)], 
with the main difference 
that $q$ now drops sharply
for small positive $1/L_{\nu q}(x_0)$ 
as opposed to 
negative $1/L_{\nu q}(x_0)$. 
For $s_{\nu}=2/5$,
$q$ equals $-2/5,3/5,\cdots$
on resonance and $|1/L_{\nu q}(x_0)| = 0$ for $x_0 \rightarrow 0$.
We 
note that the trapped $N$-body system supports a 
deep-lying bound state if the logarithmic derivative
is negative [not shown in Fig.~\ref{figLogD}(c)].

The dependence of $q$ on $[L_{\nu q}(x_0)]^{-1}$
for $s_{\nu}=1/2$ [Fig.~\ref{figLogD}(b)] is 
distinctly different from that for $s_{\nu}=3/5$ and $2/5$.
Figure~\ref{figLogD} shows that
the $q$ values depend much more weakly
on $x_0$ for $s_{\nu}=1/2$ than for $s_{\nu} \ne 1/2$.
Furthermore, $q$ approximately equals  $0,1,\cdots$ for $1/L_{\nu}(x_0) =0$.
As $|1/L_{\nu q}(x_0)|$ increases, the $q$ values change gradually
and approach half-integer values for large $|1/L_{\nu q}(x_0)|$
(on resonance, $q=-1/2,1/2,\cdots$).
In summary,
for $s_{\nu} \ne 1/2$, the system exhibits
an $N$-body resonance in
the ZR limit for $1/L_{\nu q}(x_0)=0$;
for $q=-s_{\nu}$, e.g., the third term on the right hand
side of Eq.~(\ref{eq_logD}) vanishes 
and $\lim_{x_0 \rightarrow 0}|L_{\nu q}(x_0)|=\infty$.
For $s_{\nu} = 1/2$, in contrast, the system exhibits
an $N$-body resonance in
the ZR limit for $1/L_{\nu q}(x_0)=\infty$;
for $q=-s_{\nu}$, e.g., the first and the third term on the right hand
side of Eq.~(\ref{eq_logD}) vanish 
and $\lim_{x_0 \rightarrow 0}L_{\nu q}(x_0)=0$.

We now introduce a framework that
expresses the resonance condition  
in terms of the generalized 
energy-dependent scattering length 
${\cal{V}}_{s_{\nu}}(E)$.
To define 
${\cal{V}}_{s_{\nu}}(E)$,
we consider the free-space system, i.e., we set 
the trapping frequency $\omega$  to $0$.
Compared to the trapped system, only 
the hyperradial Schr\"odinger equation 
changes [see Eqs.~(\ref{eq_poteff}) and (\ref{eq_radial})].
The hyperradial solution for the free-space system
with positive energy $E$ can be written analogously
to the solution for the trapped system
[see Eq.~(\ref{eq_capfgeneral})],
\begin{eqnarray}
\label{eq_scatteringgeneral}
F_{\nu k}(R)= \nonumber \\
N_{\nu k}
\left[  R^{1/2} J_{s_{\nu}}(kR)
- \tan(\delta_{s_\nu}(k)) R^{1/2} Y_{s_\nu }(k R) \right],
\end{eqnarray}
where we use the continuous variable $k$ to 
label the solution of the hyperradial Schr\"odinger equation;
$k$ is defined in terms of the $N$-body scattering energy $E$
and the hyperradial mass $\mu$,
$k=\sqrt{2 \mu E/\hbar^2}$.
In Eq.~(\ref{eq_scatteringgeneral}),
$J_{s_{\nu}}$ and $Y_{s_{\nu}}$ denote the Bessel 
functions of the 
first and second kind, respectively.
The phase shift $\delta_{s_{\nu}}(k)$
characterizes
the $N$-body scattering process and
can be used to define the 
generalized
energy-dependent $N$-body
scattering length ${\cal{V}}_{s_{\nu}}(k)$,
\begin{eqnarray}
{\cal{V}}_{s_{\nu}}(k) = -\frac{\tan(\delta_{s_{\nu}}(k))}{k^{2{s_{\nu}}}} 
\frac{2^{2{s_{\nu}}}\Gamma({s_{\nu}})\Gamma({s_{\nu}}+1)}{\pi};
\label{scattV}
\end{eqnarray}
throughout this paper, the generalized energy-dependent
scattering length is written, depending on the context, as
${\cal{V}}_{s_{\nu}}(k)$ or ${\cal{V}}_{s_{\nu}}(E)$.
The power of $k$ in the denominator 
on the right hand
side of Eq.~(\ref{scattV}) is chosen such 
that the generalized
energy-independent $N$-body
scattering length ${\cal{V}}_{s_{\nu}}(0)$,
${\cal{V}}_{s_{\nu}}(0) =\lim_{k \rightarrow 0} {\cal{V}}_{s_{\nu}}(k)$,
is well-behaved, i.e., 
such that ${\cal{V}}_{s_{\nu}}(k)$
approaches a constant in the $k \rightarrow 0$
limit.
The ``extra'' factors on the right hand side of
Eq.~(\ref{scattV}) are chosen such 
that ${\cal{V}}_{s_{\nu}}(0)$ reduces to
$R_0^{2s_{\nu}}$ if 
the boundary condition $F_{\nu k}(R_0)=0$ is imposed.
We note that the definition of ${\cal{V}}_{s_{\nu}}(E)$, Eq.~(\ref{scattV}),
has some similarities with 
that employed to describe the scattering of two
particles
with vanishing azimuthal quantum number
in two spatial dimensions~\cite{verhaar,kanjilal}.
For $s_{\nu}=1/2$, the $N$-body hyperradial scattering problem 
becomes formally identical to the scattering between
two three-dimensional $s$-wave interacting 
particles and 
${\cal{V}}_{s_{\nu}}(0)$ becomes formally equivalent to 
the usual three-dimensional $s$-wave scattering length $a_s$.
In general, however, the generalized scattering length
has units of (length)$^{2s_{\nu}}$.

To express the resonance condition for the trapped
system in terms of the
generalized energy-dependent
scattering length ${\cal{V}}_{s_{\nu}}(E)$, 
we relate the eigenenergies of the
trapped $N$-body system to 
${\cal{V}}_{s_{\nu}}(E)$.
To this end,
we calculate the logarithmic derivative of the 
free-space solution $F_{\nu k}(R)$,
Eq.~(\ref{eq_scatteringgeneral}), for small $x_0$.
Considering $0 < s_{\nu}< 1$ and
keeping terms up to order $x^{2s_{\nu}-1}$, we find
\begin{eqnarray}
\label{eq_logD2}
L_{\nu k}(x_0) \approx \frac{\frac{1}{2}-s_{\nu}}{x_0} - 
\frac{2^{s_{\nu}+1} s_{\nu} a_{\mathrm{ho}}^{2 s_{\nu}}}{{\cal{V}}_{s_{\nu}}(E)} x_0^{2s_{\nu}-1}
-
\nonumber \\
\frac{2^{-s_{\nu}+1} \pi \cot(\pi s_{\nu})}{[\Gamma(s_{\nu})]^2} 
\left(\frac{E}{\hbar\omega} \right)^{s_{\nu}}
x_0^{2s_{\nu}-1}.
\end{eqnarray}
For $s_{\nu}=1/2$, 
the first and third terms on the right hand side
of Eq.~(\ref{eq_logD2}) vanish and 
we find
$L_{\nu k}(x_0) \approx -\sqrt{2}a_{\mathrm{ho}}/{\cal{V}}_{s_{\nu}}(E)$ or
$L_{\nu k}(R_0) \approx -[{\cal{V}}_{s_{\nu}}(E)]^{-1}$.
To obtain an explicit relationship between the eigenenergies
of the trapped $N$-body system and the 
generalized energy-dependent
scattering length ${\cal{V}}_{s_{\nu}}(E)$
for $0 < s_{\nu}<1$,
we expand the logarithmic derivative of the trapped
system, Eq.~(\ref{eq_logD}), up to order $x_0^{2s_{\nu}-1}$
and set it equal to the
logarithmic derivative of the free-space system 
[right hand side of Eq.~(\ref{eq_logD2})],
\begin{eqnarray}
\label{eq_trans}
\frac{a_{\mathrm{ho}}^{2s_{\nu}}}{{\cal{V}}_{s_{\nu}}(E)}
+ 
h(E,s_{\nu})
=
\frac{\Gamma(\frac{-E}{2\hbar \omega} +\frac{1+s_{\nu}}{2} ) \Gamma(1-s_{\nu})}
{\Gamma(\frac{-E}{2\hbar \omega} +\frac{1-s_{\nu}}{2}) \Gamma(1+s_{\nu}) 2^{s_{\nu}}},
\end{eqnarray}
where
\begin{eqnarray}
\label{eq_energydeph}
h(E,s_{\nu})=
\frac{\pi \cot(\pi s_{\nu})}{2^{2s_{\nu}} \Gamma(s_{\nu}) \Gamma(1+s_{\nu})}
\left( \frac{E}{\hbar \omega} \right)^{s_{\nu}} .
\end{eqnarray}
It can be seen that the $1/x_0$ divergencies are canceled and, furthermore,
that Eq.~(\ref{eq_trans}) is---at this level of
approximation---independent of $x_0$.
Equation~(\ref{eq_trans}) determines 
the energy of the trapped $N$-body system
in terms of the generalized
energy-dependent scattering 
length 
${\cal{V}}_{s_{\nu}}(E)$ and can be solved self-consistently
for the eigenenergies $E$. 

For $s_{\nu}=1/2$, $h(E,s_{\nu})$
vanishes
and 
$\Gamma(1-s_{\nu})/[\Gamma(1+s_{\nu})2^{s_{\nu}}]$
reduces to $\sqrt{2}$.
Setting $[L_{\nu q}(x_0)]^{-1}$
equal to
$-{\cal{V}}_{s_{\nu}}(E)/(\sqrt{2} a_{\mathrm{ho}})$
[see discussion after Eq.~(\ref{eq_logD2})], 
Fig.~\ref{figLogD}(b)
can be interpreted as showing 
the non-integer quantum number $q$ of the
trapped $N$-body system as a function of 
$-{\cal{V}}_{s_{\nu}}(E)/(\sqrt{2}a_{\mathrm{ho}})$.
A vanishing generalized energy-dependent
scattering length implies an infinitely large logarithmic derivative,
and an infinitely large generalized energy-dependent 
scattering length implies a vanishing logarithmic derivative.
From Eq.~(\ref{eq_trans}), it follows for $s_{\nu}=1/2$ that 
$E=E_{g,\nu q}$ if ${\cal{V}}_{s_{\nu}}(E)$ diverges,
i.e., the $N$-body resonance occurs when $[{\cal{V}}_{s_{\nu}}(E)]^{-1}$
vanishes. 
The analysis outlined here for $s_{\nu}=1/2$ is formally identical to 
that of the trapped two-particle system.
Identifying ${\cal{V}}_{s_{\nu}}(E)$ with the
$s$-wave atom-atom scattering length, Eq. (\ref{eq_trans}) is identical
to
the well-known eigenequation
for two trapped $s$-wave interacting particles
with ZR interactions~\cite{busc98}.

\begin{figure}
\vspace*{1.5cm}
\includegraphics[angle=0,width=65mm]{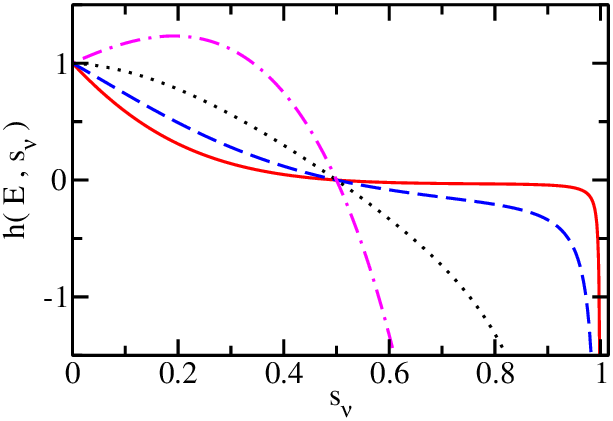}
\vspace*{-0.2cm}
\caption{(Color online)
Solid, dashed, dotted and dash-dotted 
lines show the quantity $h(E, s_{\nu})$,
Eq.~(\protect\ref{eq_energydeph}), as a function of $s_{\nu}$
for $E =1 \times 10^{-2} \hbar \omega$,
$1 \times 10^{-1} \hbar \omega$, 
$1 \hbar \omega$, and
$10 \hbar \omega$, respectively.
}\label{fig_energydeph}
\end{figure}

\begin{figure}
\vspace*{1.5cm}
\includegraphics[angle=0,width=65mm]{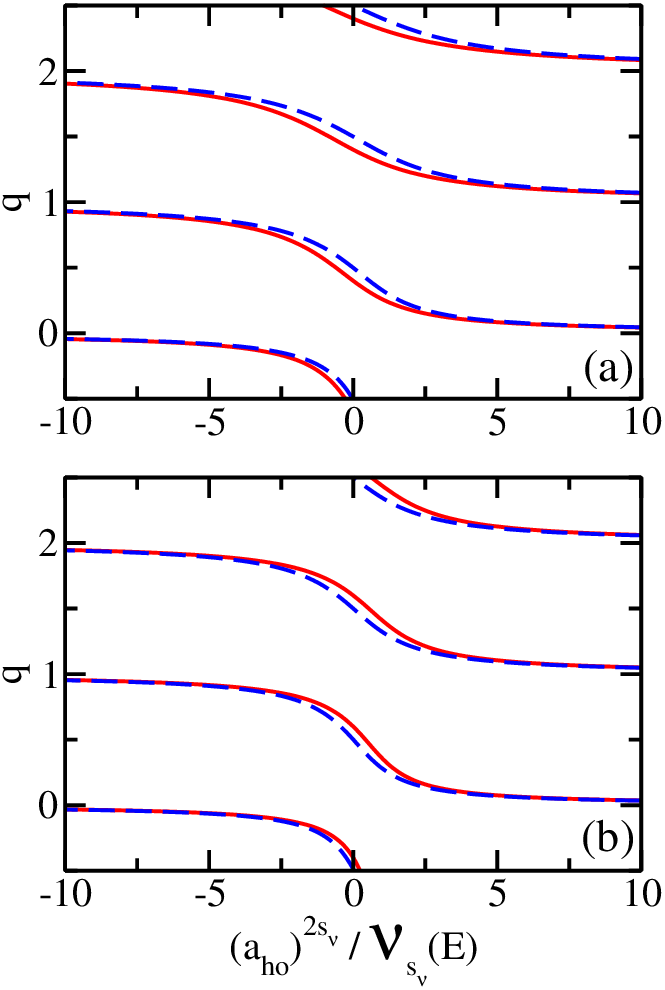}
\vspace*{-0.2cm}
\caption{(Color online)
Dashed lines show the non-integer quantum
number $q$
of
the trapped $N$-body system, 
obtained by solving Eq.~(\ref{eq_trans})
self-consistently,
as a function of 
the generalized energy-dependent
scattering length ${\cal{V}}_{s_{\nu}}(E)$
for (a) $s_{\nu}=3/5$ and (b) $s_{\nu}=2/5$. 
For comparison, solid lines show the $q$ values
that result
when $h(E,s_{\nu})$ is artificially set to zero.
}\label{figScatt}
\end{figure}

For $s_{\nu} \ne 1/2$ and $E>0$, 
$h(E,s_{\nu})$ does not
vanish 
and
introduces an additional energy dependence
on the left hand side of Eq.~(\ref{eq_trans}),
which originates from
the explicit energy-dependence of the logarithmic derivative of the
free-space solution. If we artificially set $h(E,s_{\nu})$ to zero,
the eigenenergy $E$ equals $E_{g,\nu q}$
when ${\cal{V}}_{s_{\nu}}(E)$ diverges.
Inclusion of $h(E,s_{\nu})$
shifts the energy for all ${\cal{V}}_{s_{\nu}}(E) \ne 0$
($s_{\nu} \ne 1/2$)
and modifies
the resonance condition.
In particular, the resonance condition becomes
\begin{eqnarray}
-a_{\mathrm{ho}}^{2 s_{\nu}}/{\cal{V}}_{s_{\nu}}(E_{g,\nu q})=h(E_{g,\nu q},s_{\nu}).
\end{eqnarray}
Figure~\ref{fig_energydeph} shows $h(E,s_{\nu})$
as a function of $s_{\nu}$
for four different energies, i.e., 
$E= 10^{-2} \hbar \omega- 10 \hbar \omega$.
$h(E,s_{\nu})$ vanishes if $s_{\nu}$ equals $1/2$ 
(see also above),
and increases with increasing $|s_{\nu}-1/2|$.
For $E>0$, $h(E,s_{\nu})$ takes on negative values for $1/2<s_{\nu} < 1$
and positive values for $0<s_{\nu} < 1/2$.
As can be seen, $h(E,s_{\nu})$
approaches $-\infty$ as $s_{\nu}$ approaches 1. 
Thus, for large $|{\cal{V}}_{s_{\nu}}(E)/a_{\mathrm{ho}}^{2s_{\nu}}|$ 
and $s_{\nu} \ne 1/2$, the
$a_{\mathrm{ho}}^{2 s_{\nu}}/{\cal{V}}_{s_{\nu}}(E)$ term is small and the
$h(E,s_{\nu})$ term 
dominates the left hand side
of Eq.~(\ref{eq_trans}).
For small  $|{\cal{V}}_{s_{\nu}}(E)/a_{\mathrm{ho}}^{2 s_{\nu}}|$, in contrast,
the
$a_{\mathrm{ho}}^{2 s_{\nu}}/{\cal{V}}_{s_{\nu}}(E)$ term is large and the
$h(E,s_{\nu})$ term 
on the left hand side
of Eq.~(\ref{eq_trans})
can be neglected.
Figures~\ref{figScatt}(a) and (b) show the eigenenergy of the trapped
system as a function of the generalized
energy-dependent
scattering length ${\cal{V}}_{s_{\nu}}(E)$
for $s_{\nu}=3/5$ and $2/5$, respectively.
Dashed lines show the non-integer quantum number $q$ 
obtained by 
solving Eq.~(\ref{eq_trans}) self-consistently
while solid lines 
show the non-integer quantum number $q$ that results
when 
$h(E,s_{\nu})$
is artificially set
to $0$. Figure~\ref{figScatt} shows that the inclusion of
the $h(E,s_{\nu})$ term shifts the eigenenergies
up (down) compared
to those calculated for
$h(E,s_{\nu})=0$ for $s_{\nu}=3/5$
($s_{\nu}=2/5$).
The term $h(E,s_{\nu})$ 
has a small effect when $|{\cal{V}}_{s_{\nu}}(E)|$
is small but introduces a notable shift of the
energies  when $|{\cal{V}}_{s_{\nu}}(E)|$
is large.

As discussed after Eq.~(\ref{eq_energyf}), the 
ZR framework employed in this section
does not allow for the determination of the 
actual value of the logarithmic derivative or, equivalently, the
generalized 
scattering length.
Let us imagine that the 
generalized scattering length ${\cal{V}}_{s_{\nu}}(E)$
can be controlled experimentally by tuning the interactions
between like atoms~\cite{nishida} or 
by varying the parameters of a lattice confinement~\cite{nishida}.
The realization of an $N$-body resonance
for $s_{\nu} \ne 1/2$ requires extreme fine-tuning
since both $h(E,s_{\nu})$  and ${\cal{V}}_{s_{\nu}}(E)$
vary with energy. 
On the other hand, for $s_{\nu}=1/2$, only  
${\cal{V}}_{s_{\nu}}(E)$ varies with energy. The energy-dependence of
${\cal{V}}_{s_{\nu}}(E)$ is expected
to be weakest for $s_{\nu}=1/2$ since the 
effective angular momentum barrier in the
$R$ coordinate vanishes in this case.
This reasoning is motivated by what is known from
``ordinary'' two-particle 
scattering (see, e.g., Ref.~\cite{newtonscattering}):
In general, two-particle $s$-wave
scattering (no angular
momentum barrier) exhibits a much weaker energy-dependence
than two-particle $p$- or $d$-wave scattering
(finite angular momentum barrier).
We conclude that it should be more likely to
realize an $N$-body resonance if $s_{\nu}=1/2$ than 
if $s_{\nu} \ne 1/2$.

Lastly,
we note that application of the two-body scattering framework 
to the hyperradial problem at hand 
implies that
the free-space system supports an $N$-body
bound state with zero energy when 
${\cal{V}}_{s_{\nu}}(0)$ diverges.
Furthermore,
the $N$-body system supports a 
single weakly-bound free-space bound
state when ${\cal{V}}_{s_{\nu}}(E)$ is large
and positive,
and no 
weakly-bound 
free-space bound state when ${\cal{V}}_{s_{\nu}}(E)$ is negative.

The following section introduces the stochastic variational
(SV)
approach which we use to numerically solve the Schr\"odinger equation
in the relative coordinates
for few-body systems interacting through FR potentials.
This approach employs Jacobi coordinates
instead of hyperspherical coordinates and 
makes no assumption about the small $R$ behavior
of $F_{\nu q}(R)$: the small $R$
behavior is not treated as input but instead is 
a natural part of the solution.
Connections between the results based
on the approaches discussed in
Secs.~\ref{sec_hyperspherical}
and \ref{sec_sv} will
be made in Sec.~\ref{sec_results}.

\subsection{Stochastic variational treatment}
\label{sec_sv}
The SV approach~\cite{cgbook,sore05,stec07}
 expands the relative
wave function $\Psi$ 
in terms of  basis functions $\varphi_k$,
\begin{eqnarray}
\Psi=\sum_{k=1}^{N_{\mathrm{basis}}} c_k {\cal{A}} \left[ \varphi_k(\vec{x})
\right],
\end{eqnarray}
where the $c_k$ denote expansion coefficients,
${\cal{A}}$ denotes an operator that ensures the anti-symmetry
of the basis functions, and $\vec{x}$ collectively
denotes the Jacobi vectors $\vec{\rho}_j$, where $j=1,\cdots,N-1$.
The basis functions 
[see Eqs.~(\ref{eq_cgbasis}) and (\ref{eq_cgbasis2}) 
for their explicit forms]
are chosen such that the Hamiltonian matrix can be constructed analytically.
The eigenenergies $E$ of the relative
Hamiltonian,
which provide an upper bound to the
exact eigenenergies, are then obtained by
diagonalizing the Hamiltonian matrix.
The variational bound can be improved systematically 
by increasing the size of the basis set (i.e., by increasing $N_{\mathrm{basis}}$)
and by varying the parameters $\vec{u}^{(k)}$ and $d_{ij}^{(k)}$
[see Eq.~(\ref{eq_cgbasis})]
or
$\vec{s}^{(k)}$ and $d_{ij}^{(k)}$ [see Eq.~(\ref{eq_cgbasis2})].
Our primary interest in this work is to describe the energetically
lowest-lying state of various $(N_1,N_2)$ systems.
The energetically lowest lying state of the $(2,1)$ system, e.g.,
has natural parity (see Fig.~\ref{fig_s0_n3})
while that of the $(3,1)$ system
has unnatural parity. Thus, we need to consider basis functions that
can describe both natural and unnatural parity states.

To treat natural parity states, i.e., states
with $\Pi=(-1)^L$,
we employ basis functions that are written in terms of the 
spherical harmonic $Y_{L0}$, which
determines the relative orbital angular
momentum of the system and
depends on the 
unit vector $\hat{v}^{(k)}$~\cite{cgbook}, 
\begin{eqnarray}
\label{eq_cgbasis}
\varphi_k(\vec{x}) =
|\vec{v}^{(k)}|^{L} {Y}_{L0}
(\hat{v}^{(k)})
\exp \left[-
\sum_{i<j}^N \left(\frac{r_{ij}}{\sqrt{2}d_{ij}^{(k)}} \right)^2
\right],
\end{eqnarray}
where
\begin{eqnarray}
\vec{v}^{(k)} = \sum_{j=1}^{N-1} u_j^{(k)} \vec{\rho}_j.
\end{eqnarray}
Here, the $u_j^{(k)}$
form
a $(N-1)$-dimensional parameter vector
that
determines how the relative orbital angular momentum $L$
of the $(N_1,N_2)$ system
is distributed among the $(N-1)$
Jacobi vectors $\vec{\rho}_j$.
We find that 
the optimal set of widths $d_{ij}^{(k)}$ 
depends quite strongly on the mass ratio $\kappa$.
For the $(2,1)$ system with equal-masses, e.g., three-body bound states
are absent~\cite{skor57,petrov}. 
This implies that three-body correlations are largely
absent, and that the contribution of
basis functions with more than one
$d_{ij}^{(k)}$ of the order of the range of the underlying two-body
potential contribute negligibly to the wave function~\cite{stec07,stec08}.
When $\kappa \approx 12$, in contrast, three-body correlations 
are non-negligible and
basis functions $\varphi_k$ that are characterized by 
three widths $d_{ij}^{(k)}$ of the order of the range of the 
two-body potential contribute notably~\cite{blum10}.

To describe states with unnatural parity, we employ so-called
geminal-type
basis functions $\varphi_k$ that are neither eigenfunctions
of the angular momentum operator nor the parity 
operator~\cite{cgbook,dail10},
\begin{eqnarray}
\label{eq_cgbasis2}
\varphi_k(\vec{x})=
\exp \left[-
\sum_{i<j}^N \left(\frac{r_{ij}}{\sqrt{2}d_{ij}^{(k)}} \right)^2 +
(\vec{s}^{(k)})^T \vec{x} \right].
\end{eqnarray}
Here,
$(\vec{s}^{(k)})^T \vec{x}$ is just the dot product
between two $3(N-1)$ dimensional vectors.
The $3(N-1)$-dimensional parameter vector
$\vec{s}^{(k)}$ is, together with the
$N(N-1)/2$ widths $d_{ij}^{(k)}$, optimized 
semi-stochastically for each basis function $\varphi_k$. 
In general, the determination of the natural parity states
[see Eq.~(\ref{eq_cgbasis})] 
is, for the same $N$, numerically significantly more
efficient than that of unnatural parity states
[see Eq.~(\ref{eq_cgbasis2})]. 

The SV approach also allows for the determination of 
structural properties. To calculate structural properties, we follow two
different approaches~\cite{blum10a}: 
(i) We implement the analytically
known matrix elements~\cite{cgbook} 
for the operator $A$ of interest, and
calculate the expectation value of $A$ using the known 
$c_k$, where $k=1,\cdots,N_{\mathrm{basis}}$.
(ii) We calculate the expectation value of the operator
$A$ by sampling the normalized density $|\psi|^2$ through 
Metropolis sampling~\cite{mcbook}.
In the limit of infinitely many Monte Carlo samples, the 
results of approach (ii) should agree with those of approach (i).
Where possible, we have used this to check our implementations.  
To calculate the hyperradial density $P_{\mathrm{hyper}}(R)$,
where 
\begin{eqnarray}
\int_0^{\infty} P_{\mathrm{hyper}}(R) dR=1,
\end{eqnarray}
we employ approach (ii). The
pair distribution functions
$P_{\mathrm{hl}}(r)$ and $P_{\mathrm{hh}}(r)$
for heavy-light and heavy-heavy atom pairs, normalized such that
\begin{eqnarray}
4 \pi \int_0^{\infty} P_{\mathrm{hl}}(r) r^2 dr = 1
\end{eqnarray}
[and similarly for $P_{\mathrm{hh}}(r)$], 
is calculated using approach (i) for
the wave function written in terms of the basis
functions given in Eq.~(\ref{eq_cgbasis})
and approach (ii) for that 
written in terms of the basis
functions given in Eq.~(\ref{eq_cgbasis2}).

\section{Results}
\label{sec_results}
This section presents our results from the SV calculations
for FR interactions
and interprets our findings within the hyperspherical framework.
Sections~\ref{sec_threebody}
and \ref{sec_fourbody} present
the energetics and structural properties
for the $(2,1)$ and $(3,1)$ systems, respectively.

\subsection{Three-body resonances}
\label{sec_threebody}
\begin{figure}
\vspace*{1.5cm}
\includegraphics[angle=0,width=65mm]{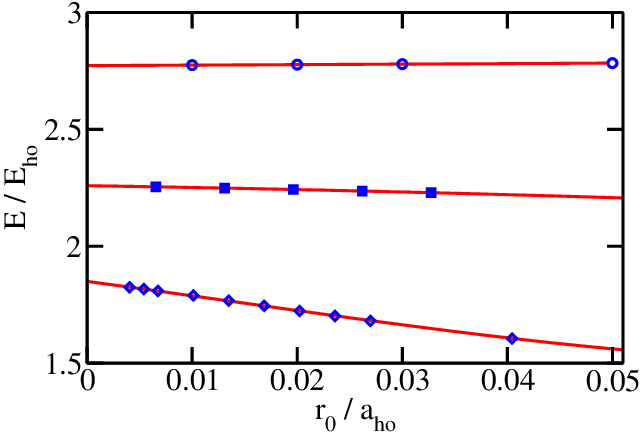}
\vspace*{0.0cm}
\caption{(Color online)
Circles, squares and diamonds show the SV energies for the
$(2,1)$ system with $L^{\Pi}=1^-$
at unitarity for $\kappa=1, 6$ and 10, respectively.
The unlike particles interact through $V_{\mathrm{g}}$.
Solid lines show three-parameter fits 
of the form $\sum_{j=0}^2 c_j r_0^j$ to the SV energies.
}\label{fig_n3_energy_small}
\end{figure}

\begin{figure}
\vspace*{1.5cm}
\includegraphics[angle=0,width=65mm]{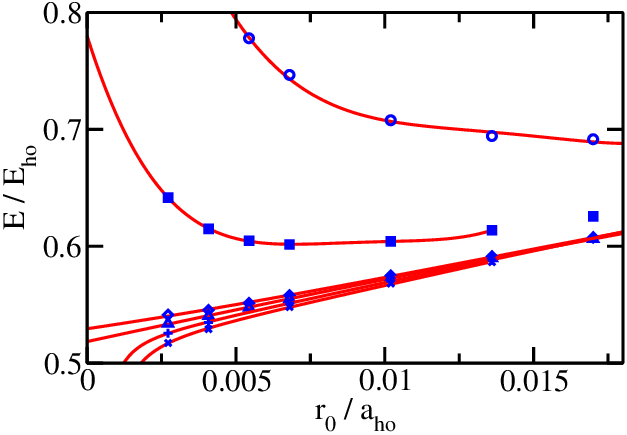}
\vspace*{0.0cm}
\caption{(Color online)
Circles, squares, diamonds, triangles, pluses and 
crosses show the SV energies for the
$(2,1)$ system with $L^{\Pi}=1^-$
at unitarity for $\kappa=12.25, 12.3, 12.3131, 12.314, 12.315$ 
and $12.316$, respectively.
The unlike particles interact through $V_{\mathrm{g}}$.
Solid lines show three- to five-parameter fits to the SV energies.
}\label{fig_n3_energy_large}
\end{figure}

We start our discussion with the infinitely strongly
interacting $(2,1)$  system
at unitarity, and then discuss the behavior
away from the two-body resonance.
Since the $L^{\Pi}=1^-$ state
is, as discussed in Sec.~\ref{sec_hyperspherical}, 
the energetically lowest lying state for all $\kappa$ of interest,
we restrict our
SV calculations to this symmetry.
For each mass ratio $\kappa$, we consider different 
$r_0$ with $r_0 \ll a_{\mathrm{ho}}$. 
Symbols in Figs.~\ref{fig_n3_energy_small}
and \ref{fig_n3_energy_large} show examples for 
the ground state energy of 
systems with $\kappa=1$ through $12.316$ interacting through $V_{\mathrm{g}}$.
By fitting the range-dependent SV energies 
to simple 
three- to five-parameter
expressions, we extrapolate the FR SV energies
to the $r_0 \rightarrow 0$ limit.
For $\kappa=1$ and $6$
(circles and squares in Fig.~\ref{fig_n3_energy_small}), 
the SV energies approach 
the 
ZR limit approximately linearly from above and below, respectively. 
For $\kappa=10$, 
the FR energies are best described by a quadratic three-parameter fit.
As $\kappa$ increases further,
the range-dependence increases notably
(see
circles and squares in Fig.~\ref{fig_n3_energy_large} 
for $\kappa=12.25$ and $12.3$). 
For $\kappa = 12.3131$ and $12.314$ (diamonds and triangles
in Fig.~\ref{fig_n3_energy_large}),
in contrast, the 
range-dependence is comparatively small and the SV energies 
approach the ZR limit
approximately
linearly from above.
Finally, for $\kappa=12.316$ (crosses in Fig.~\ref{fig_n3_energy_large}) 
the energies decrease as $1/r_0^2$
for sufficiently small $r_0$, indicating the
presence of a three-body bound state.
For yet larger $\kappa$ (not shown in
Fig.~\ref{fig_n3_energy_large}),
the energy of the trapped system becomes negative.
The appearance of the three-body bound state
indicates the presence of a three-body resonance.
We estimate the resonance for the potential
$V_{\mathrm{g}}$ in the $r_0 \rightarrow 0$ limit
to be located at $\kappa \approx 12.314(2)$.
The uncertainty in our estimate for the resonance position
arises mainly from the extrapolation of
our SV energies to the ZR limit.
The SV energies for $\kappa=12.315$, e.g., are nearly
equally well
described by fits of the form $c_{-2}/r_0^2+c_{-1}/r_0 +c_0$
(shown in Fig.~\ref{fig_n3_energy_large})
and
$c_0+c_1r_0+c_2r_0^2$ 
(not shown in Fig.~\ref{fig_n3_energy_large}).

\begin{figure}
\vspace*{1.5cm}
\includegraphics[angle=0,width=65mm]{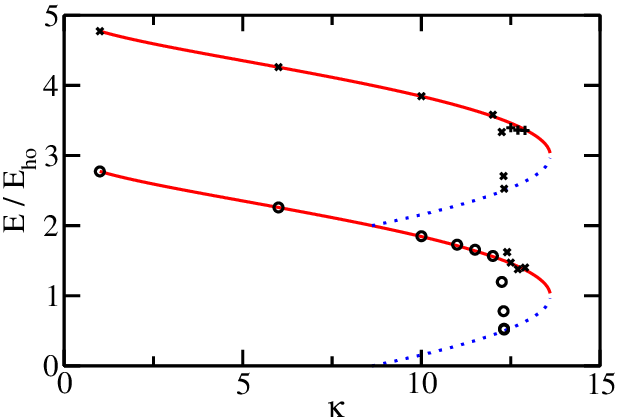}
\vspace*{0.0cm}
\caption{(Color online)
Energies of the $(2,1)$ system with $L^{\Pi}=1^-$ at unitarity.
Circles, crosses and pluses show the SV energies
for $V_{\mathrm{g}}$,
extrapolated to the ZR limit, for the three
energetically lowest-lying states  as a function of $\kappa$.
Solid and dotted lines show
$E_{f,0q}$ ($q=0$ and $1$) and $E_{g,0q}$ ($q=-s_0$ and $-s_0+1$),
respectively. This figure has been adapted from Ref.~\cite{blum10}.
}\label{fig_energyn3}
\end{figure}

Symbols
in Fig.~\ref{fig_energyn3} show the extrapolated ZR energies
for the three energetically lowest lying states
of the $(2,1)$ system with $L^{\Pi}=1^-$ 
interacting through
$V_{\mathrm{g}}$ with $1/a_s=0$.
The dropping of the energies around $\kappa \approx 12.3$
associated with the three-body resonance
is clearly visible: The ground state energy (circles)
becomes negative while the
energies of the first excited state (crosses) 
and the second excited state (pluses) drops by
approximately $2 \hbar \omega$.
For comparison, solid and dotted lines in Fig.~\ref{fig_energyn3}
show the ZR energies $E_{f,0q}$ [Eq.~(\ref{eq_energyf}) with $q=0$ and $1$]
and $E_{g,0q}$ 
[Eq.~(\ref{eq_energyg}) with $q=-s_0$ and $-s_0+1$], respectively.
Away from the three-body resonance, the extrapolated ZR energies 
agree well with $E_{f,0q}$. On resonance, the extrapolated ZR energies agree
well with $E_{g,0q}$.
Our analysis is fully consistent with the
general discussion of Ref.~\cite{castin}, 
where it was found that universal states for systems under spherically 
symmetric harmonic confinement
must have an energy larger than $1 \hbar \omega$ and that
states with energy less than $1 \hbar \omega$ are necessarily
non-universal.
The admixture of the irregular solution $g_{\nu q}$
requires that the boundary condition
of the hyperradial wave function be specified, 
which makes the system properties 
dependent on an additional
parameter and thus non-universal.
It has been pointed out previously~\cite{castin} 
that
the three-body system is again scale-invariant
at the three-body resonance.
Our work (see also Ref.~\cite{blum10}) 
shows an example for this non-universal
three-body resonance. This non-trivial three-body resonance 
has also recently been
investigated by Gandolfi and Carlson~\cite{gand10}, who
studied the free-space problem.

The mass ratio at which the three-body resonance exists
depends on the details of the underlying two-body
potential.
As has been argued in Ref.~\cite{blum10}
and in Sec.~\ref{sec_hyperspherical},
it seems most likely that the three-body resonance
occurs when the effective hyperangular momentum
barrier vanishes, i.e., when $s_{\nu}\approx 1/2$.
For the $(2,1)$ system with ZR interactions and $L^{\Pi}=1^-$
symmetry,
this is the case when $\kappa \approx 12.3131$
(see Fig.~\ref{fig_s0_n3});
this value is close to the resonance position found for the 
purely attractive Gaussian potential $V_{\mathrm{g}}$.
In our example, the trimer on resonance is
large 
[this follows from Eq.~(\ref{eq_radial_compact})];
this implies a long lifetime
and thus opens the intriguing
possibility of studying trimer correlations
in many-body
systems.

\begin{figure}
\vspace*{1.5cm}
\includegraphics[angle=0,width=65mm]{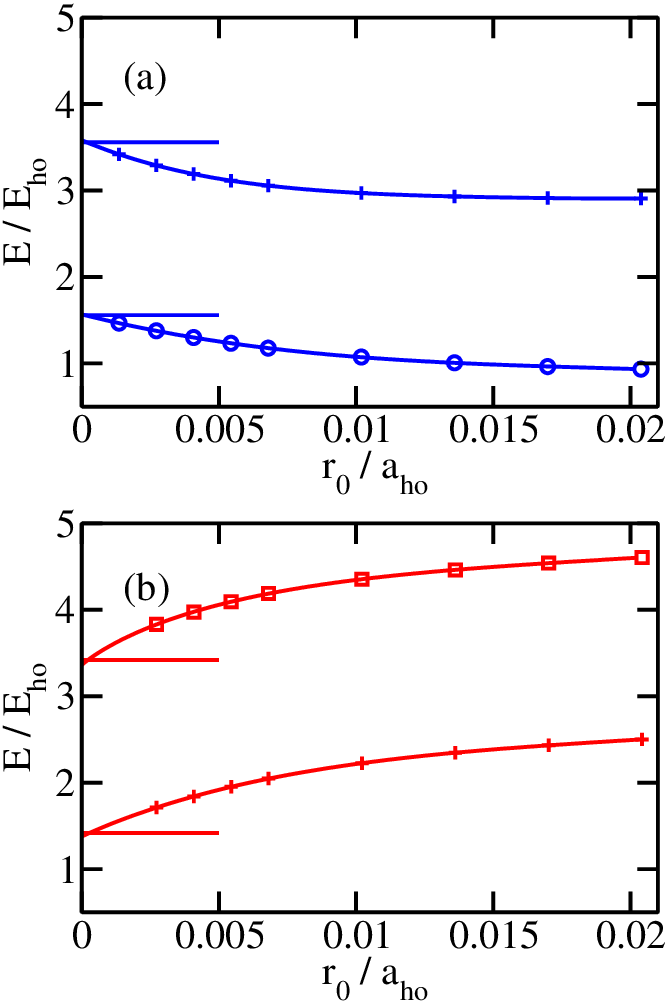}
\vspace*{0.0cm}
\caption{(Color online)
Symbols show the SV energies for 
the $(2,1)$ system with $L^{\Pi}=1^-$
interacting through $V_{\mathrm{g}}$
at unitarity for
(a) $\kappa=12$ and
(b) $\kappa=12.7$, respectively,
as a function of $r_0$.
For $\kappa=12$, the energies of the ground state (circles)
and first excited
state (pluses) are shown.
For $\kappa=12.7$, the energies of the first and second excited
states (pluses and squares) are shown.
Solid lines show fits to the SV energies.
Horizontal lines show the energy $E_{f,0q}$ for $q=0$ and
$1$.
}\label{fig_n3_excited}
\end{figure}

Symbols in Figs.~\ref{fig_n3_excited}(a) and (b) shows the SV energies
for the $(2,1)$ system with $\kappa=12$ and $12.7$, respectively,
interacting through $V_{\mathrm{g}}$ with $1/a_s=0$ 
as a function of $r_0$.
For both $\kappa$,
the 
SV energies for the ``lowest universal
state'' and the ``second lowest universal
state'' are shown
[for $\kappa=12.7$, the ground state (not shown)
corresponds to a tightly bound
non-universal trimer].
The extrapolated ZR energies of the two universal
states agree well with the energies $E_{f,0q}$ ($q=0$ and $1$),
which are shown by horizontal lines.
This implies that the 
$2 \hbar \omega$ spacing expected 
for universal states~\cite{castin} is
fullfilled with good accuracy.
For $\kappa=12$ and $12.7$, our
fits
result in an energy spacing of
$2.014 \hbar \omega$ and $1.984 \hbar \omega$, respectively.
For finite $r_0$, the spacing between the two lowest
universal states deviates from the $2 \hbar \omega$ spacing.
For $r_0 \approx 0.01a_{\mathrm{ho}}$, e.g., we find a spacing of 
$\approx 1.90 \hbar \omega$ and $2.13 \hbar \omega$ for
$\kappa=12$ and $12.7$, respectively.

\begin{figure}
\vspace*{1.5cm}
\includegraphics[angle=0,width=65mm]{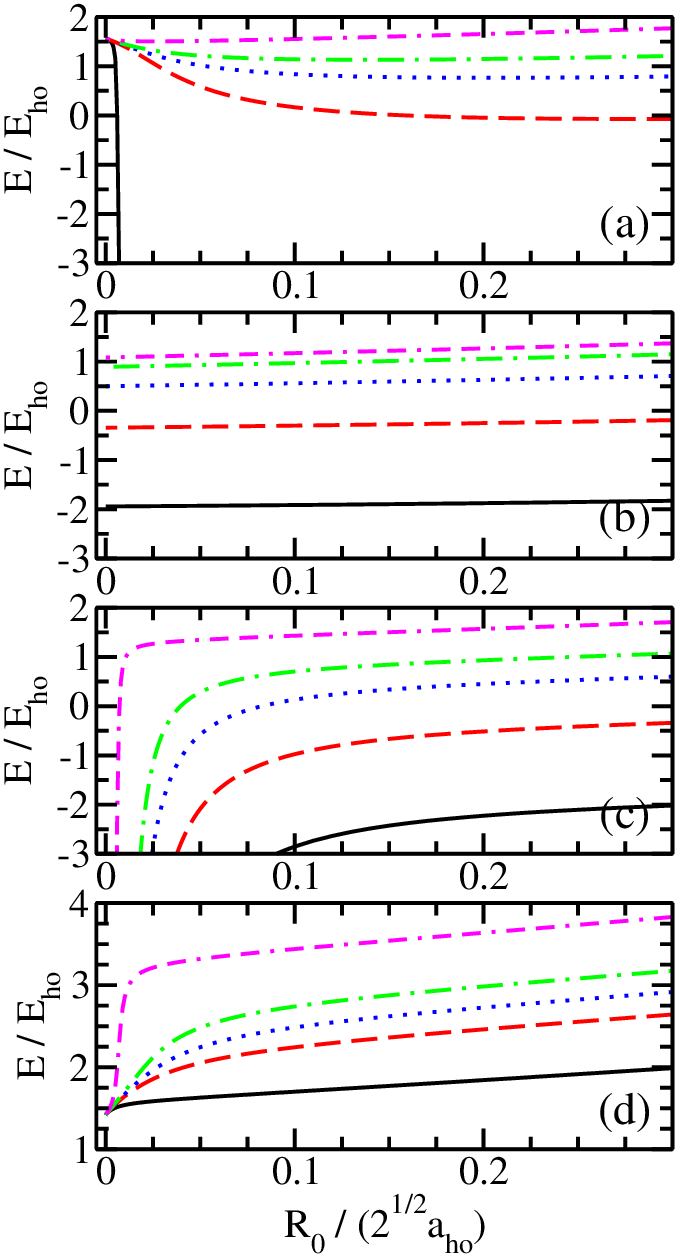}
\vspace*{-0.2cm}
\caption{
(Color online) 
Eigenenergy $E$ of the trapped
system at unitarity, obtained from Eq.~(\ref{eq_logD}), 
as a function of $R_0$ for
(a) $s_{\nu} \approx 0.5579$ (lowest allowed $q$ value),
(b) $s_{\nu} =1/2$ (lowest allowed $q$ value),
(c) $s_{\nu} \approx 0.4180$ (lowest allowed $q$ value), and
(d) $s_{\nu} \approx 0.4180$ (second lowest allowed $q$ value).
Dash-dash-dotted, dash-dotted, dotted, dashed and solid lines
correspond to 
(a) $L_{\nu q}(x_0)=10, 1, 0, -1,$ and $-10$,
(b) $L_{\nu q}(x_0)=2, 1, 0, -1,$ and $-2$,
(c) $L_{\nu q}(x_0)=10, 1, 0, -1,$ and $-2$, and
(d) $L_{\nu q}(x_0)=10, 1, 0, -1,$ and $-10$.
}\label{fig_logderivativex0}
\end{figure}

To make an explicit connection between the FR SV energies and the 
energies obtained within the hyperspherical framework
(see Sec.~\ref{sec_hyperspherical}),
we assume a direct proportionality between the
range $r_0$ of the two-body interaction potential
and the hyperradius $R_0$ at which the logarithmic
derivative of the hyperradial function $F_{\nu q}(R)$
is imposed. The proportionality factor between $r_0$ and $R_0$ 
cannot be determined within the model;
we find that a proportionality factor of
order $5$ to $10$, i.e., $r_0\approx 5-10 R_0$, is appropriate.
Figures~\ref{fig_logderivativex0}(a)-(d) exemplarily show the
eigenenergies predicted by the hyperspherical framework
as a function of $R_0$ for $s_{\nu}\approx0.5579$ and the lowest allowed $q$ value,
for $s_\nu=1/2$ and the lowest allowed $q$ value, 
and for $s_{\nu}=0.4180$ 
and the lowest and second lowest allowed $q$ values
for various values of the
logarithmic derivative.
For the $(2,1)$ system 
with $L^{\Pi}=1^-$ symmetry, these $s_{\nu}$ values
correspond to $\kappa=12$, $12.3131$ and $12.7$, respectively.
We find that
there exists a value of the
logarithmic derivative for each $\kappa$ that predicts energies 
as a function of $R_0$ [via Eq.~(\ref{eq_logD})] that are in qualitative
agreement with the dependence of the SV energies on $r_0$.
For example, the SV energies of the lowest universal 
state for $\kappa=12.7$ approach the ZR limit from above
[pluses in Fig.~\ref{fig_n3_excited}(b)], 
in agreement with the dependence of the energies
obtained within the hyperspherical framework 
on $R_0$ [see Fig.~\ref{fig_logderivativex0}(d)].
For $\kappa=12$, in contrast, the SV energies of the ground
state approach the ZR limit from below
[circles in Fig.~\ref{fig_n3_excited}(a)],
in agreement with the dependence of the energies
obtained within the hyperspherical framework
on $R_0$ for small $|L_{\nu q}|$ 
[see Fig.~\ref{fig_logderivativex0}(a)].
Although the comparison between the SV energies and those
obtained within the hyperspherical framework leads to
a consistent picture (including the existence
or absence of bound states), 
the analysis unfortunately does not
allow for the unambiguous extraction of the
value of the logarithmic derivative or the generalized scattering 
length.

\begin{figure}
\vspace*{1.5cm}
\includegraphics[angle=0,width=60mm]{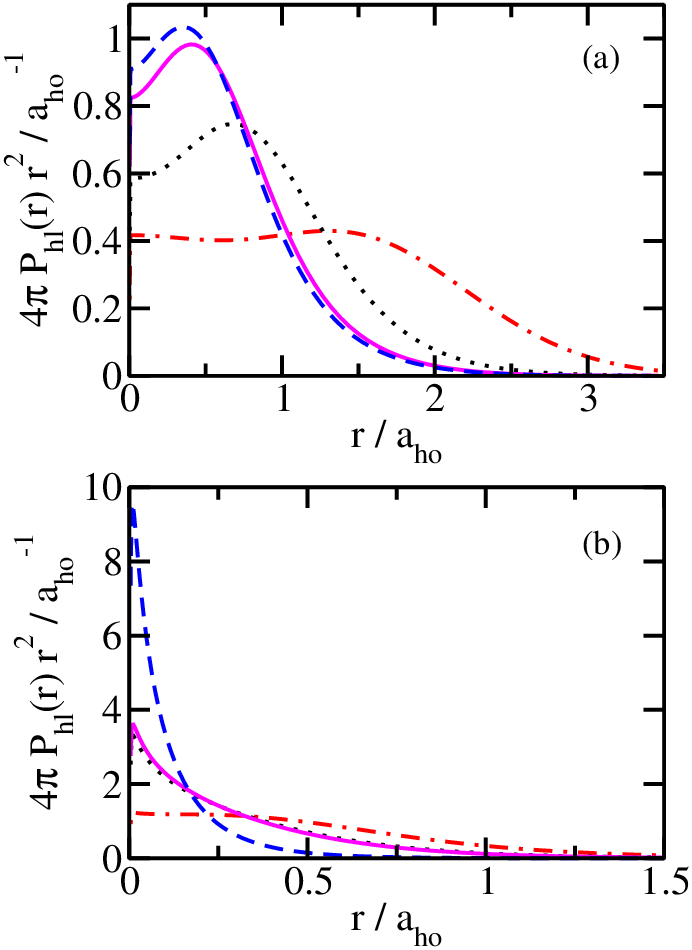}
\vspace*{0.0cm}
\caption{(Color online)
Dash-dotted, dotted, solid, and dashed 
lines show the scaled pair 
distribution function $P_{\mathrm{hl}}(r)r^2$ for the 
heavy-light pair of the
$(2,1)$ system at unitary with $L^{\Pi}=1^-$
for (a) $\kappa=1, 6.7, 11$, and $11.5$, 
and (b) $\kappa=12, 12.3, 12.314$, and $12.5$, respectively.
The heavy-light particles interact through $V_{\mathrm{g}}(r)$
with $r_0=0.003a_{\mathrm{ho}}$.
Note the different scales of the axis in panels (a) and (b).
}\label{fig_n3_pair_updown}
\end{figure}

\begin{figure}
\vspace*{1.5cm}
\includegraphics[angle=0,width=60mm]{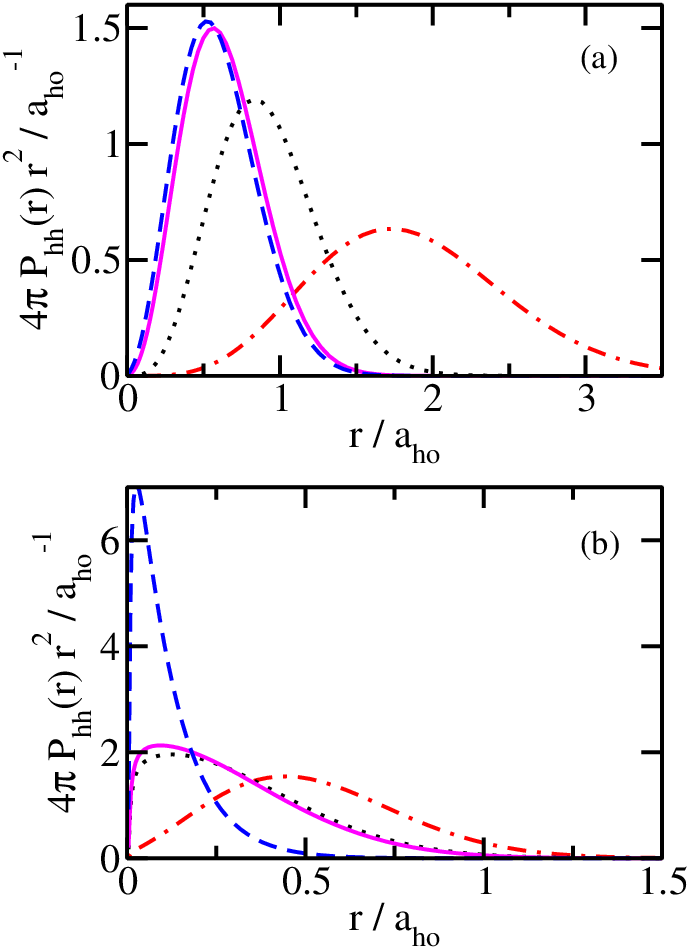}
\vspace*{0.0cm}
\caption{(Color online)
Dash-dotted, dotted, solid, and dashed 
lines show the scaled pair 
distribution function $P_{\mathrm{hh}}(r)r^2$ for the 
heavy-heavy pair of the
$(2,1)$ system at unitary with $L^{\Pi}=1^-$
for (a) $\kappa=1, 6.7, 11$, and $11.5$, and
(b) $\kappa=12, 12.3, 12.314$, and $12.5$, respectively.
The heavy-light particles interact through $V_{\mathrm{g}}(r)$
with $r_0=0.003a_{\mathrm{ho}}$.
Note the different scales of the axis in panels (a) and (b).
}\label{fig_n3_pair_upup}
\end{figure}

To gain additional insight into the 
three-particle system,
we analyze the structural properties
of the $(2,1)$ system interacting through $V_{\mathrm{g}}$ with
$L^{\Pi}=1^-$, $1/a_s=0$ and $r_0=0.003 a_{\mathrm{ho}}$.
Figure~\ref{fig_n3_pair_updown}
shows the scaled pair distribution function $P_{\mathrm{hl}}(r)r^2$
for the heavy-light pairs for $\kappa=1$ through $12.5$,
while Fig.~\ref{fig_n3_pair_upup}
shows the scaled pair distribution function $P_{\mathrm{hh}}(r)r^2$
for the heavy-heavy pair
for the same mass ratios.
As $\kappa$ increases, the amplitudes of 
$P_{\mathrm{hl}}(r)r^2$ and $P_{\mathrm{hh}}(r)r^2$
increase at small distances
and decrease at large distances.
The scaled 
pair distribution function
for the spin-up---spin-down distance coordinate~\cite{footnoteupdown} for
$\kappa=1$ [dash-dotted lines
in Fig.~\ref{fig_n3_pair_updown}(a)], e.g., shows a ``two-bump
structure'' that has been 
previously interpreted within an atom-dimer
picture~\cite{stec08}: 
The ``bump'' at smaller $r$, $r \approx 0.1 a_{\mathrm{ho}}$, 
reflects the formation of a dimer while
the ``bump'' at 
larger $r$, $r \approx 1.5a_{\mathrm{ho}}$, reflects the fact that the
``spare'' spin-up atom sits further away from the spin-down atom
than the spin-up atom that forms the dimer.
For $\kappa=6.7$ [dotted lines in
Figs.~\ref{fig_n3_pair_updown}(a) and
\ref{fig_n3_pair_upup}(a)], which corresponds to the
$^{40}$K-$^6$Li mixture, the 
heavy-light and heavy-heavy pair distribution functions
reflect the effective attraction
between the two heavy atoms. Compared to the $\kappa=1$ system,
the likelihood of finding the two like
atoms at distances smaller than $a_{\mathrm{ho}}$ is
significantly increased.
For $\kappa=12.5$, both the scaled heavy-light and
heavy-heavy pair distribution functions fall
off approximately exponentially
at length scales smaller than $a_{\mathrm{ho}}$, as expected for
a three-body bound state.

\begin{figure}
\vspace*{1.5cm}
\includegraphics[angle=0,width=60mm]{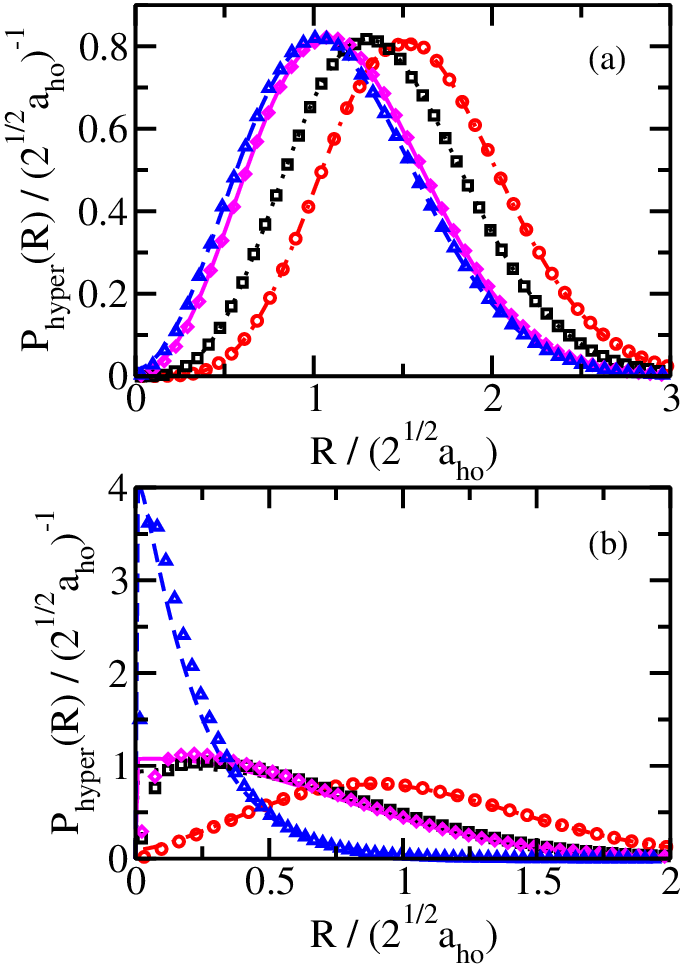}
\vspace*{0.0cm}
\caption{(Color online)
Hyperradial density $P_{\mathrm{hyper}}(R)$
of the
$(2,1)$ system 
with $L^{\Pi}=1^-$ at unitary.
Symbols show the results from the SV calculation
where the heavy-light particles interact through $V_{\mathrm{g}}(r)$
with $r_0=0.003a_{\mathrm{ho}}$
for (a) $\kappa=1$ (circles), $\kappa=6.7$ (squares), $\kappa=11$ 
(diamonds), and $\kappa=11.5$ (triangles), and
(b) $\kappa=12$ (circles), $\kappa=12.3$ (squares), $\kappa=12.314$ 
(diamonds), and $\kappa=12.5$ (triangles).
Lines show the hyperradial
density $|F_{0q}(x)|^2$, Eq.~(\ref{eq_radial_compact}).
The $s_0$ entering into Eq.~(\ref{eq_radial_compact}) is 
taken
from the ZR model.
$q$ is set to $0$ for $\kappa=1$ and $6.7$,
and calculated according
to Eq.~(\ref{eq_energy}) with $E$ taken from the SV calculations
for the other $\kappa$.
Note the different scales of the axis in panels (a) and (b).
}\label{fig_n3_hyper}
\end{figure}

Symbols in Fig.~\ref{fig_n3_hyper}
show the hyperradial density $P_{\mathrm{hyper}}(R)$
for the FR interaction $V_{\mathrm{g}}$
with $r_0=0.003a_{\mathrm{ho}}$ and $1/a_s=0$
for various $\kappa$.
For comparison, lines show the hyperradial densities
$|F_{\nu q}(R)|^2$
obtained from the ZR model.
As can be seen from Eq.~(\ref{eq_radial_compact}), 
$F_{\nu q}(R)$ depends on $s_{\nu}$ and $q$. For all $\kappa$ considered
in Fig.~\ref{fig_n3_hyper}, we use the $s_0$ obtained by solving
the hyperangular Schr\"odinger equation for ZR
interactions. The quantum number $q$ is set to zero
for $\kappa=1$ and $6.7$ [see Eq.~(\ref{eq_energyf})].
For $\kappa>8.619$,
we adjust the short-range boundary condition of $F_{\nu q}(R)$
so as to reproduce the FR energies, i.e., we
calculate $q$ according to Eq.~(\ref{eq_energy}) 
with $E$ taken from the FR SV calculation.
Figure~\ref{fig_n3_hyper}
shows good agreement between the hyperradial densities
obtained from the SV calculations and those
obtained within the ZR model.
For $\kappa=12$, e.g., 
the admixture of the irregular solution $g_{0 q}$
is clearly reflected in the hyperradial density:
Using $q=0$,
i.e., using $f_{0 q}$ only, results in a notably poorer description
of the system (not shown).

\begin{figure}
\vspace*{1.5cm}
\includegraphics[angle=0,width=65mm]{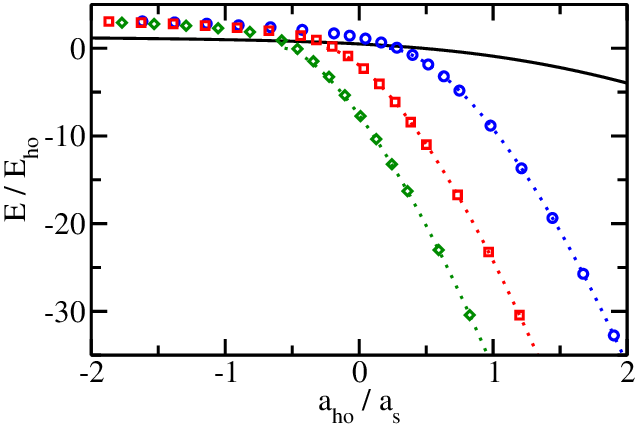}
\vspace*{0.0cm}
\caption{(Color online)
Circles, squares and diamonds show the
SV energies for the $(2,1)$ system with $L^{\Pi}=1^-$
interacting through $V_{\mathrm{g}}$ with $r_0=0.005a_{\mathrm{ho}}$
as a function of $a_{\mathrm{ho}}/a_s$ for $\kappa=12$, 
$12.314$ and $12.5$, respectively.
Dotted lines show fits to the SV energies (see text for details).
For comparison, the solid line shows the energy for two
trapped atoms interacting through $V_{\mathrm{zr}}$.
}\label{fig_n3_energy_as}
\end{figure}

So far, we have considered systems with infinitely large
two-body $s$-wave scattering length.
To shed further light on the $(2,1)$ system
in the vicinity of the
three-body resonance, 
we also performed calculations for finite $a_s$.
For the free-space $(2,1)$ system with $L^{\Pi}=1^-$ and
positive $a_s$, 
a single universal three-body bound state whose energy is a few times 
that of the two-body system has been predicted to exist
for $\kappa \approx 8.173$ to $12.917$,
and two universal bound states 
have been predicted to exist for $\kappa \approx 12.917$ to
$13.606$~\cite{russian}.
The three-body bound states discussed here for the trapped 
system are distinctly different from these universal states,
i.e., they do not approach those discussed in Ref.~\cite{russian}
when $\omega$ approaches 0.
Symbols in Fig.~\ref{fig_n3_energy_as}
show 
the energies of the trapped three-body system
as a function of $1/a_s$ for the interaction potential
$V_{\mathrm{g}}$ with $r_0=0.005a_{\mathrm{ho}}$
for three different mass ratios $\kappa$,
$\kappa=12$ (below the three-body resonance),
$\kappa=12.314$ (at the three-body resonance), and
$\kappa=12.5$ (above the three-body resonance).
We note that the three-body energies depend strongly
on $r_0$. 
As expected, for negative (positive) $a_s$, the three-body energy
lies above (below) the corresponding energy at unitarity.
For comparison,
a solid line shows the energy of the trapped
two-body system with ZR interactions. For small $|a_s|$ ($a_s>0$),
the two-body energy varies approximately as $1/a_s^2$.
To obtain a semi-quantitative description
of the three-body energies on the positive 
scattering length side, we 
write
$E/E_{\mathrm{ho}} = c_1 (a_{\mathrm{ho}}/a_s-c_3)^{c_2}$.
Fitting our three-body energies for $a_{\mathrm{ho}}/a_s>0.5$,
we find that
the $c_2$ coefficient changes from $1.65$ over $1.55$ to $1.52$
as $\kappa$ changes from $12$ over $12.314$ to $12.5$.
The resulting fits are shown by
dotted lines in Fig.~\ref{fig_n3_energy_as}.

As discussed in Ref.~\cite{blum10},
we also considered systems consisting of two heavy and 
two light fermions
interacting through $V_{\mathrm{g}}$ with $1/a_s=0$ and various angular
momenta. 
The addition of the light particle to the $(2,1)$ system does not, 
to within our numerical resolution, lead to the
appearance of a new resonance. Thus, adding a light particle leaves the
system properties largely unchanged. The next subsection shows
that adding a heavy fermion to the $(2,1)$
system does lead to the appearance of a new resonance.

\subsection{Four-body resonances}
\label{sec_fourbody}
Unlike the hyperangular Schr\"odinger equation 
for the $(2,1)$ system with ZR interactions, that for the
$(3,1)$ system with ZR interactions is not analytically soluble.
Thus, 
we employ the SV approach 
to solve the full Schr\"odinger equation
for FR interactions,
and then analyze the eigenenergies 
and structural properties. In certain cases,
we ``back out'' $s_0$ and $q$, i.e., 
we extract the lowest eigenvalue of the hyperangular 
Schr\"odinger equation and the eigenvalues
of the radial Schr\"odinger equation.

\begin{figure}
\vspace*{1.5cm}
\includegraphics[angle=0,width=65mm]{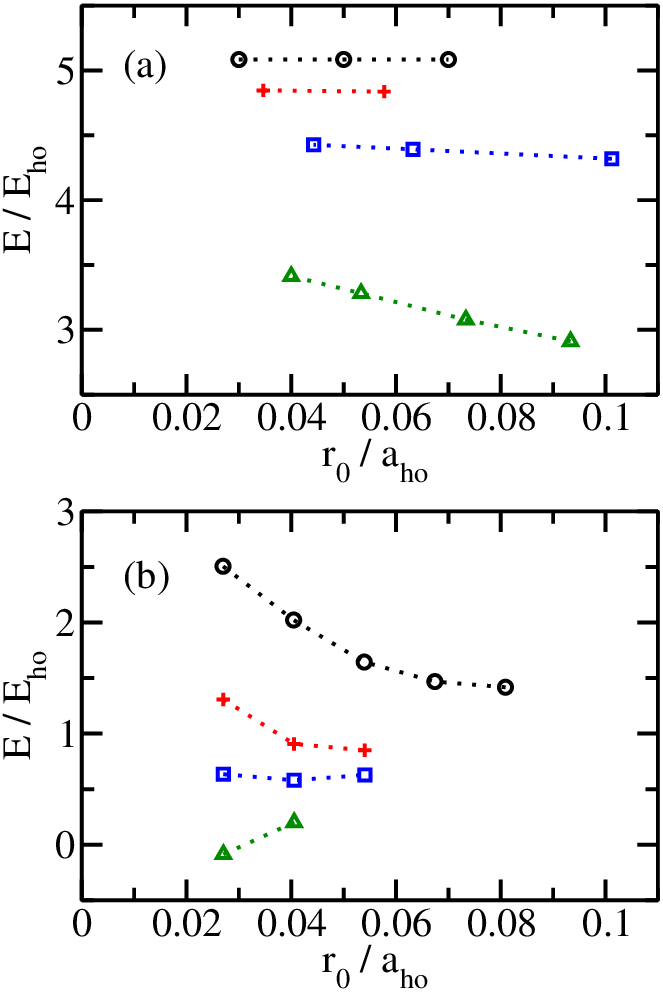}
\vspace*{0.0cm}
\caption{(Color online)
Circles, pluses, squares and triangles show the
SV energies for the $(3,1)$ system with $L^{\Pi}=1^+$
interacting through $V_{\mathrm{g}}$ at unitarity
as a function of $r_0$ for (a) 
$\kappa = 1,2,4$ and $8$,
and (b) 
$\kappa = 10, 10.4, 10.5$ and $10.6$.
Dotted lines 
are shown as a guide to the eye.
}\label{fig_energy31_r0}
\end{figure}

\begin{figure}
\vspace*{1.5cm}
\includegraphics[angle=0,width=65mm]{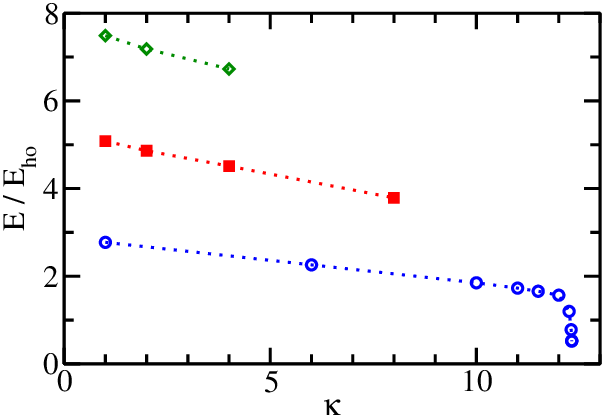}
\vspace*{0.0cm}
\caption{(Color online)
Circles, squares and diamonds show the extapolated ZR
energies as a function of $\kappa$ for
the $(2,1)$ system with $L^{\Pi}=1^-$ symmetry,
the $(3,1)$ system with $L^{\Pi}=1^+$ symmetry, and
the $(4,1)$ system with $L^{\Pi}=0^-$ symmetry, respectively.
As a guide to the eye,
dotted lines connect data points.
For the $(3,1)$ and $(4,1)$ systems, our FR SV energies 
allow for a reliable extrapolation to the $r_0 \rightarrow 0$
limit only for relatively small $\kappa$.
}\label{fig_energy31}
\end{figure}

The energetically lowest lying state of the $(3,1)$ system has 
$L^{\Pi}=1^+$ symmetry for the 
two-body scattering lengths $a_s$
of interest.
In the $a_s\rightarrow 0^-$ limit, this can be 
verified by constructing the wave functions 
of the non-interacting system. We find numerically that this is also
true at unitarity.
Figures~\ref{fig_energy31_r0}(a) and (b)
show the energy of the $(3,1)$
system as a function of $r_0$ for different $\kappa$.
Similarly to the $(2,1)$ system, 
the extrapolated ZR energy of the $(3,1)$ system
is approached approximately linearly from above
for $\kappa=1$. For larger $\kappa$ (i.e., $\kappa=4$ or 8), 
the ZR limit is
approached approximately linearly from below.
Table~\ref{tab_energy31} 
\begin{table}
\caption{Extrapolated ZR $s_0$ values for 
small two-component Fermi gases at unitarity.
The uncertainty of the $s_0$ values is
estimated to be in the last digit reported.
The $(2,2)$ energies for $\kappa=1$
are taken from Ref.~\protect\cite{dail10}
and the $(3,2)$ energy for $\kappa=1$ 
is taken from Ref.~\protect\cite{blum10a}.}
\begin{ruledtabular}
\begin{tabular}{c|cccccc}
$\kappa$ & $s_0(3,1)$ & $s_0(4,1)$ & $s_0(2,2)$ & $s_0(2,2)$ & $s_0(2,2)$ & $s_0(3,2)$ \\
 & $1^+$ & $0^-$ & $0^+$ & $1^-$ & $2^+$ & $1^-$ \\ 
\hline
1  & 4.08 & 6.45 & 2.509 & 4.598 & 3.418 & 4.958  \\
2  & 3.86 & 6.15 & 2.575 & 4.357 &       & 4.90  \\
4  & 3.51 & 5.68 & 2.754 & 3.997 & 3.478 & 4.85  \\
6  &      &      & 2.886 & 3.705 &       & 4.76  \\
8  & 2.79 &      & 2.947 & 3.430 & 3.326 &   \\
10 &      &      & 2.939 & 3.138 & 3.225 &   
\end{tabular}
\end{ruledtabular}
\label{tab_energy31}
\end{table}
and squares in
Fig.~\ref{fig_energy31} summarize the extrapolated ZR energies
for the $(3,1)$ system with $L^{\Pi}=1^+$ symmetry for $\kappa \le 8$.
The dependence of the energies on the
range $r_0$ increases as $\kappa$ increases from $1$ or
$2$ to about $10$,
and then becomes comparatively small 
for $\kappa \approx 10.5$
[see squares in Fig.~\ref{fig_energy31_r0}(b)].
For yet larger $\kappa$, the energy varies roughly
as $1/r_0^2$ and approaches large negative values 
for small $r_0$. In this regime, four-body bound states exist.
Although our calculations for larger $\kappa$, $\kappa > 8$,
do not allow for a reliable extrapolation of the energies
to the $r_0 \rightarrow 0$ limit,
the existence of a four-body or $(3,1)$ resonance
around $\kappa \approx 10.4$ or $10.5$ is evident.

Although our SV calculations for the $(3,1)$ system are restricted to
significantly larger $r_0$ than those for the $(2,1)$ system,
it is clear that
the behavior of the $(3,1)$ energies 
in the vicinity of the four-body resonance
is qualitatively similar to that of the $(2,1)$ energies 
in the vicinity of
the three-body resonance.
The main difference is that the four-body resonance occurs
at a smaller mass ratio than the three-body resonance.
Importantly, the SV energies for the Gaussian two-body potential
$V_{\mathrm{g}}$ for the $(3,1)$ system depend---as do those for the $(2,1)$ 
system---comparatively weakly on $r_0$ in the vicinity
of the 
resonance [squares in Fig.~\ref{fig_energy31_r0}(b)].
The qualitatively similar
dependence of the energies of the $(3,1)$ and $(2,1)$ systems
on $r_0$ suggests that the four-body resonance 
occurs when $s_0$ is approximately equal to $1/2$.
More specifically, applying the framework detailed in
Sec.~\ref{sec_hyperspherical}, i.e.,
assuming that the hyperradial and the hyperangular
motion separate not only for ZR interactions but
also
for FR interactions, the relatively weak dependence
of the $(3,1)$ energies for $\kappa \approx 10.5$ 
implies that the four-body resonance 
occurs when $s_0$ is approximately $1/2$.
This interpretation assumes separability or 
approximate separability
of the hyperangular and hyperradial parts of the wave function.
If the coupling between different channel functions
was appreciable for the FR interactions considered in our 
SV calculations,
then the hyperspherical framework of Sec.~\ref{sec_hyperspherical}
would need to be modified. While it is possible that 
the dependence of our SV energies on $r_0$
is consistent with a value of $s_0$ notably different from
$1/2$ and non-vanishing channel
coupling, we believe that the former scenario 
(i.e., approximate separability and $s_0 \approx 1/2$ 
in the vicinity of the
four-body resonance) is more likely.
This conclusion is supported by our analysis of the hyperradial
densities.

\begin{figure}
\vspace*{1.5cm}
\includegraphics[angle=0,width=65mm]{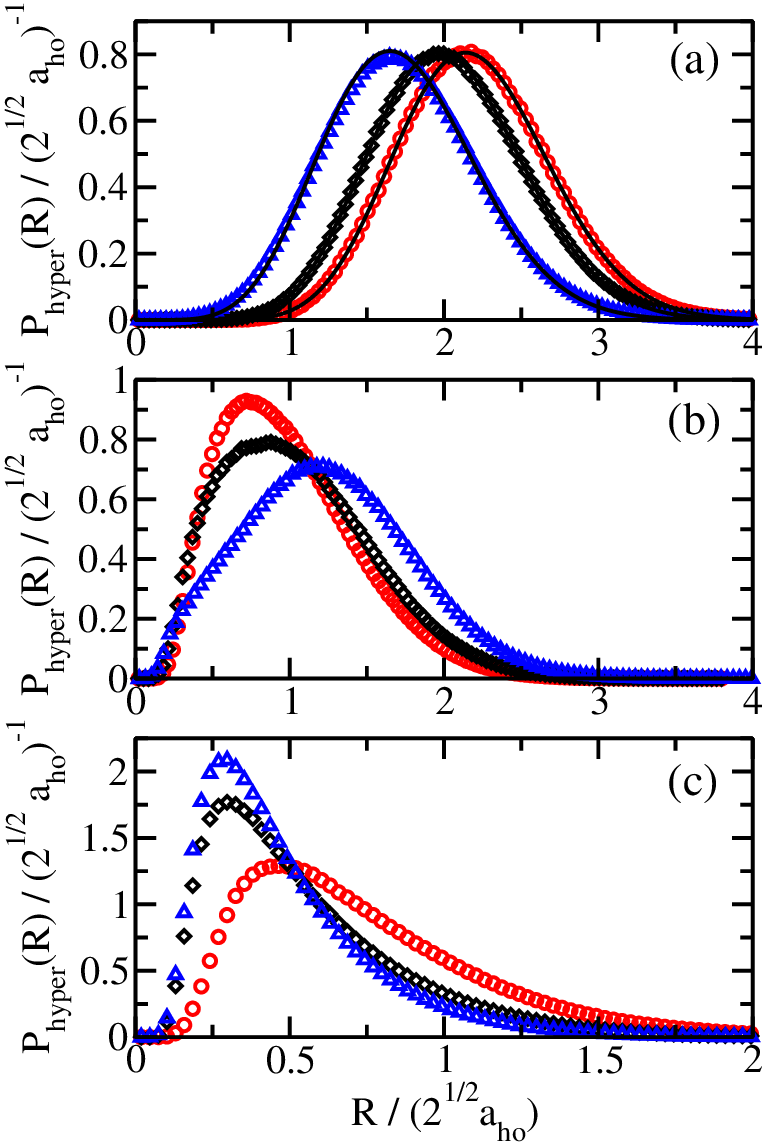}
\vspace*{0.0cm}
\caption{(Color online)
Hyperradial density $P_{\mathrm{hyper}}(R)$
of the
$(3,1)$ system 
with $L^{\Pi}=1^+$ at unitary.
Symbols show the results from the SV calculation
where the heavy-light particles interact through $V_{\mathrm{g}}(r)$
for 
(a) 
$\kappa=1$ and $r_0=0.03 a_{\mathrm{ho}}$ (circles), 
$\kappa=4$ and $r_0 \approx 0.044 a_{\mathrm{ho}}$ (diamonds), and 
$\kappa=8$ and $r_0= 0.04 a_{\mathrm{ho}}$ (triangles);
(b) 
$\kappa=10$ and $r_0 \approx 0.054 a_{\mathrm{ho}}$ (circles), 
$\kappa=10$ and $r_0 \approx 0.040 a_{\mathrm{ho}}$ (diamonds), and 
$\kappa=10$ and $r_0 \approx 0.027 a_{\mathrm{ho}}$ (triangles);
and
(c) 
$\kappa=10.4$ and $r_0 \approx 0.041 a_{\mathrm{ho}}$ (circles), 
$\kappa=10.5$ and $r_0 \approx 0.027 a_{\mathrm{ho}}$ (diamonds), and 
$\kappa=10.6$ and $r_0 \approx 0.027 a_{\mathrm{ho}}$ (triangles).
In panel~(a), lines show the hyperradial
density $|F_{0q}(R)|^2$, Eq.~(\ref{eq_radial_compact}),
with $q$ set to zero and $s_0$ determined by fitting the SV density
to Eq.~(\ref{eq_radial_compact}).
Note the different scales of the axis in panels (a)-(c).
}\label{fig_hyper31}
\end{figure}

Symbols in Fig.~\ref{fig_hyper31}
show the hyperradial density $P_{\mathrm{hyper}}(R)$ 
for the $(3,1)$ system with $\kappa=1-10.6$
interacting through
the FR two-body potential $V_{\mathrm{g}}$ with various $r_0$.
For $\kappa \le 8$, we find that the hyperradial densities
for the Gaussian potential $V_{\mathrm{g}}$ (symbols) are well reproduced by
$|F_{0q}(R)|^2$,
Eq.~(\ref{eq_radial_compact}), with $q$ set to zero
and $s_0$ determined by fitting
the SV densities to Eq.~(\ref{eq_radial_compact}) 
[solid
lines in Fig.~\ref{fig_hyper31}(a)].
Alternatively,
we set $q=0$ and use the
SV energies to determine $s_0$ via Eq.~(\ref{eq_energyf}).
For $\kappa=1$ and $4$, 
the resulting hyperradial densities are indistinguishable from the
lines shown Fig.~\ref{fig_hyper31}(a). For $\kappa=8$, however, the 
description that treats $s_0$ as a fitting parameter leads to a 
better description. 
Symbols in Fig.~\ref{fig_hyper31}(b)
show the hyperradial densities for $\kappa=10$
and three different $r_0$ values. As $r_0$ decreases, the 
the hyperradial density moves to larger $R$ values
and approaches the hyperradial density expected for universal states.
Symbols in Fig.~\ref{fig_hyper31}(c) show the hyperradial densities for
$\kappa=10.4, 10.5$ and $10.6$. As $\kappa$ increases, 
the maximum of $P_{\mathrm{hyper}}(R)$  increases and moves to smaller
$R$ values. Moreover, the tail of $P_{\mathrm{hyper}}(R)$
for $\kappa=10.4-10.6$ starts to resemble that of a bound state.
The ZR model, Eq.~(\ref{eq_radial_compact}),
qualitatively but not quantitatively 
reproduces the FR hyperradial densities
for $\kappa=10-10.6$ if the
normalization constant $N_{0 q}$, $q$ and $s_0$ are treated
as fitting parameters. 
Based on a detailed analysis of the hyperradial densities of the 
$(2,1)$ system for various $r_0$, we believe that the ZR model
would reproduce the FR hyperradial densities of the $(3,1)$
system quantitatively
if $r_0$ was smaller.

Our calculations suggest that the $s_0$ for the $(3,1)$ system with
$L^{\Pi}=1^+$ symmetry changes notably over a small
range of mass ratios. In particular, 
we find $s_0 \approx 1/2$ for $\kappa=10.4$ or $10.5$.
Recently, Castin and coworkers~\cite{cast10} reported that
$s_0$ becomes zero for $\kappa=13.384$.
While our $s_0$ values are not in direct contradiction with
this finding, 
an extrapolation of our results suggests
that the $s_0$ value of the $(3,1)$ system goes to zero
at a smaller mass ratio than that found by Castin and coworkers.
Future studies need to investigate this question in more
detail.

We also treated the energetically lowest-lying state of the 
$(4,1)$ system at unitarity, which 
has $L^{\Pi}=0^-$ symmetry.
For this system, our calculations are restricted to the universal regime.
Diamonds in Fig.~\ref{fig_energy31} show the extrapolated ZR
energies and Table~\ref{tab_energy31} shows the
corresponding $s_0$ values.
We find that the hyperradial densities 
for the FR interaction potential (not shown) are well reproduced
by Eq.~(\ref{eq_radial_compact}) with $q=0$ and $s_0$ 
determined through a fit or by the SV energy.
For completeness, Table~\ref{tab_energy31} shows selected
$s_0$ values for the $(3,1)$, $(2,2)$ and $(3,2)$
systems.

\section{Conclusions}
\label{sec_conclusion}
We considered small fermionic two-component systems 
under external spherically symmetric
confinement with 
$1 \le \kappa \lesssim 13.607$, 
i.e., in regimes where three-body Efimov physics is absent.
Our calculations employed two different 
interaction potentials between the heavy and light atoms,
a FR and a ZR interaction potential, and 
focused primarily 
on the
unitary regime where the
interspecies $s$-wave scattering length $a_s$
becomes infinitely large. Like atoms were assumed to be non-interacting.
To address questions
Q1-Q5 (see Sec.~\ref{sec_introduction}),
we performed SV calculations for few-fermion
systems that interact through the FR interaction
potential $V_{\mathrm{g}}$, and analyzed 
a subset of the eigenspectrum and selected structural properties
such as the hyperradial density
as a function of $r_0$.
The numerical results were interpreted within
a hyperspherical framework that becomes exact in the ZR
limit.
We defined the generalized 
energy-dependent scattering length ${\cal{V}}_{s_{\nu}}(E)$,
which characterizes the 
$N$-body system in free space.
This generalized scattering length is
related to the hypervolume introduced in Ref.~\cite{tan08}
to characterize bosonic three-particle systems and can be
interpreted as being the result of an effective $N$-body
force that acts at small (or vanishing) hyperradii $R$.
We used the generalized scattering length to
connect the properties of the trapped
$N$-body system with those of the free-space system.

We found that the $(2,1)$ and $(3,1)$ systems 
at unitarity interacting through $V_{\mathrm{g}}$ exhibit 
three-body and four-body resonances when $\kappa \approx
12.314$ and $\kappa \approx 10.4$, respectively.
For the $(2,1)$ system with $L^{\Pi}=1^-$ symmetry, 
this mass ratio corresponds to,
as has been shown by an independent calculation~\cite{russian},
an $s_0$ value of approximately $1/2$.
For the $(3,1)$ system with $L^{\Pi}=1^+$ symmetry, 
the $s_0$ value is not known 
independently. We argued that the $s_0$ value
of the $(3,1)$ system 
is
approximately $1/2$ for around $\kappa=10.4$.
It would be extremely valuable if this 
could be confirmed by an independent calculation that
solves the hyperangular Schr\"odinger equation directly 
and does not rely, as our analysis, on backing $s_0$ out
from the full solution.
The three- and four-body resonances discussed here are 
obtained for the interaction potential $V_{\mathrm{g}}$.
While the occurance of $N$-body resonances does depend,
in general, on the
details  of the underlying two-body potential,
we believe that the results presented here also apply to other
short-range interaction potentials.
In particular, we argued that $N$-body resonances occur
most likely when the $s_0$ value that characterizes the 
$N$-body problem is approximately $1/2$. 
Experimentally, the value of $s_0$ can possibly be tuned by
varying the lattice confinement or the intraspecies 
interactions~\cite{nishida}.
While our calculations for the $(4,1)$ system were restricted
to $\kappa$ values for which the system behaves universally,
i.e., for which $N$-body resonances are absent,
recent work by Gandolfi and Carlson~\cite{gand10} found that the $(4,1)$
system 
in free-space
exhibits an $N$-body resonance for $\kappa \approx 9.5-9.8$.
So far, $N$-body resonances have not been observed for larger
systems and
future research needs to address whether or not such resonances 
exist. 
Our results for the $(2,1)$ and $(3,1)$
systems show,
in agreement with Ref.~\cite{gand10}, 
that the energy of the $N$-body bound state
changes more rapidly for the $(3,1)$ system than for the $(2,1)$ 
system for the same change $\Delta \kappa$ of the mass ratio.
If $N$-body resonances exist for larger
systems, this trend is expected to continue.  

Our calculations show that FR effects become increasingly
more important as the mass ratio $\kappa$ increases.
The dependence on the range of the underlying two-body potential
signals the breakdown of universality and has important implications
for theory and experiment. Since FR effects 
can be appreciable,
theoretical treatments of unequal-mass Fermi systems
based on ZR interactions may not be, in general, sufficient. 
Our calculations suggest that an analysis of
experimental results for few- or many-fermion systems
with sufficiently large mass ratio
needs to account for non-universal FR effects.
Near an $N$-body resonance, the system's lifetime will
be reduced due to the formation of
molecules. If a two-component Fermi gas is 
prepared in a regime where $N$-body bound states exist,
collapse is expected to set in.
While this is, in many instances, an unwanted phenomenon,
careful tuning in the vicinity of an $N$-body resonance
might
open the possibility to study novel physics.
On resonance, the $N$-body clusters have a size
comparable to that of
$s$-wave dimers.
At the $(2,1)$ resonance, e.g., the trimer is large. 
While bound tetramers
and pentamers (and presumably larger clusters) exist, 
their wave functions have
negligible overlap with that of the trimer. This suggests
that many-body systems with competing dimer and trimer 
interactions have finite lifetimes and may thus be
prepared experimentally.
Similarly, many-body systems with competing two-body and four-body
interactions should be experimentally accessible.
The study of these many-body systems 
is expected to uncover
novel
physics.

\section{Acknowledgements}
Support by the NSF through
grant PHY-0855332
and the ARO
as well as 
fruitful 
discussions with
S. Tan
are gratefully acknowledged.


\begin{thebibliography}{10}
\bibitem{footnotenorm1}
This can be seen by considering the small $x$-behavior of $|g_{\nu q}(x)|^2$
for $s_{\nu}=1$,
$|g_{\nu q}(x)|^2 \rightarrow 1/x$ for $s_{\nu}=1$.
Integrating $|g_{\nu q}(x)|^2$ over $x$ from $x=0$ to $x = \epsilon$
gives $\ln x |_0^{\epsilon} + C$ (where $C$ is a constant),
which diverges.
Thus, the wave function for $s_{\nu} \ge 1$
is not normalizable unless $g_{\nu q}(x)$
is eliminated.
\bibitem{footnotenorm}
If $s_{\nu}+q$ or $q$ are equal to an integer,
Eq.~(\ref{eq_radial_compact_norm}) contains, as written, infinities.
In these cases, $N_{\nu q}$ reduces to
$N_{\nu q}=\sqrt{2/[\Gamma(q+1)\Gamma(q+1+s_{\nu})]}$.
\bibitem{footnoteupdown}
For equal masses, 
the labels ``heavy'' and ``light'' 
are replaced by ``spin-up'' and ``spin-down''.
\end{thebibliography}
\end{document}